\documentclass{aa}
\usepackage{graphicx}
\usepackage{txfonts}
\usepackage{lscape} 
\newcommand{\Prot}{P_{\rm rot}}
\newcommand{\Ppeak}{P_{\rm peak}}
\newcommand{\hpeak}{h_{\rm peak}}
\newcommand{\Rvar}{R_{\rm var}}
\newcommand{\Teff}{T_{\rm eff}}
\newcommand{\logg}{\log\,g}
\newcommand{\nperiods}{32,387} 
\newcommand{\nstars}{29,860} 
\newcommand{\nstarstwo}{13,345} 
\begin{document}
\title{Stellar rotation periods from K2 Campaigns 0--18}
\subtitle{Evidence for rotation period bimodality and simultaneous variability decrease}

\author{Timo Reinhold\inst{1}, 
	Saskia Hekker\inst{1,2}}

\offprints{T. Reinhold,\email{reinhold@mps.mpg.de}}

\institute{
  \inst{1}
  Max-Planck-Institut f\"ur Sonnensystemforschung,  
  Justus-von-Liebig-Weg 3, 37077 G\"ottingen, Germany \\
  \inst{2}
  Stellar Astrophysics Centre, Department of Physics and Astronomy, 
  Aarhus University, 120 Ny Munkegade, Building 1520, DK-8000 Aarhus C, Denmark
  }
  
\date{Received day month year / Accepted day month year}

\abstract
{
Rotation period measurements of stars observed with the \textit{Kepler} mission have revealed a lack of stars at intermediate rotation periods, accompanied by a decrease of photometric variability. Whether this so-called dearth region is a peculiarity of stars in the \textit{Kepler} field, or reflects a general manifestation of stellar magnetic activity, is still under debate. The K2 mission has the potential to unravel this mystery by measuring stellar rotation and photometric variability along different fields in the sky.
}
{
Our goal is to measure stellar rotation periods and photometric variabilities for tens of thousands of K2 stars, located in different fields along the ecliptic plane, to shed light on the relation between stellar rotation and photometric variability.
}
{
We use Lomb-Scargle periodograms, auto-correlation and wavelet functions to 
determine consistent rotation periods. Stellar brightness variability is assessed by computing the variability range, $\Rvar$, from the light curve. We further apply Gaussian mixture models to search for bimodality in the rotation period distribution.
}
{
Combining measurements from all K2 campaigns, we detect rotation periods in \nstars~stars. The reliability of these periods was estimated from stars observed more than once. We find that 75--90\% of the stars show period deviation smaller than 20\% between different campaigns, depending on the peak height threshold in the periodograms. For effective temperatures below 6000\,K, the variability range shows a local minimum at different periods, consistent with an isochrone age of $\sim$750\,Myr. Additionally, the rotation period distribution shows evidence for bimodality, although the dearth region in the K2 data is less pronounced compared to the \textit{Kepler} field. The period at the dip of the bimodal distribution shows good agreement with the period at the local variability minimum.
}
{
We conclude that the rotation period bimodality is present in different fields of the sky, and is hence a general manifestation of stellar magnetic activity. The reduced variability in the dearth region is interpreted as a cancelation between dark spots and bright faculae. Our results strongly advocate that the role of faculae has been underestimated so far, suggesting a more complex dependence of the brightness variability on the rotation period.
}
\maketitle
\section{Introduction}
Owing to four years of high-precision photometry from the \textit{Kepler} space telescope, stellar rotation periods have been measured for thousands of stars \citep{McQuillan2013a, McQuillan2013b, Reinhold2013, Walkowicz2013, Nielsen2013, McQuillan2014, doNascimento2014, Garcia2014, Reinhold2015, Ceillier2016}. One particularly interesting observation was the detection of a bimodality in the rotation period distribution of the \textit{Kepler} M~dwarfs \citep{McQuillan2013a}. It was shown that the rotation period bimodality persists for hotter stars up to $\sim$5000\,K \citep{Reinhold2013, McQuillan2014, Reinhold2015}. Recently, \citet{Davenport2017} showed that the bimodality extends to even hotter stars with effective temperatures in the range $5000-6500$\,K. 

It has been proposed that this bimodality originates from two stellar populations of different ages \citep{McQuillan2013a, McQuillan2014}. This explanation is supported by the observation that the bimodality correlates with Galactic height \citep{Davenport2018}, which is assumed to scale with stellar age. These authors find that the gap of the bimodal distribution coincides with a gyrochrone age of $\sim$600\,Myr, and that the bimodality can only be seen for stars out to $\sim$525\,pc, suggesting a burst of star formation within the last 600\,Myr in the solar neighborhood.
An alternative explanation was provided by \citet{Reinhold2019} who showed that the lack of stars at intermediate rotation periods is accompanied by a decrease in photometric variability. These authors explained the low variability in these stars by a cancelation of dark spots and bright faculae, leading to a non-detection of periodicity, and therefore causing the period bimodality.

The K2 mission provides the opportunity to study the close connection between stellar rotation and photometric variability along different fields in the ecliptic plane. Apart from the period bimodality, the stellar rotation period is a fundamental quantity that can be used to infer stellar ages via gyrochronology \citep{Barnes2003,Barnes2007,Barnes2010}. Luckily, several open clusters (such as the Pleiades, Hyades, Praesepe etc.) have been observed by K2. Since all stars in a cluster are assumed to be coeval, these open clusters form an ideal test bed for the angular momentum evolution in late-type stars. Consequently, the high-precision long-term K2 observations promoted rotation period measurements for the Pleiades \citep{Rebull2016a,Rebull2016b,Stauffer2016a}, the Hyades \citep{Douglas2016,Douglas2019}, and Praesepe \citep{Rebull2017,Douglas2017,Douglas2019}, and more measurements of other open clusters and associations are currently underway.

These cluster rotation periods can now be compared to previously determined gyrochronology relations because the cluster ages are (assumed to be) known.  Interestingly, recent studies of open clusters with ages up to 1\,Gyr \citep{Curtis2019,Douglas2019} indicate that the standard formalism of gyrochronology needs to be adjusted. Stars in these clusters spin down more slowly than predicted from gyrochronology. Moreover, these authors found that the spin-down efficacy strongly depends on spectral type, with the tendency to decrease towards later spectral types. In this context, M~dwarfs represent the most extreme cases, whose rotation periods do not seem to change at all between ages of roughly 600\,Myr to 1\,Gyr. These results strongly support the hypothesis of an epoch of stalled spin-down in late-type stars, as proposed e.g. by \citet{vanSaders2016} and \citet{Metcalfe2017}. As a consequence, gyrochronology ages seem to be much more uncertain for mid-G and later spectral types. 

In this study we analyze K2 data from Campaigns 0--18 covering different fields in the sky. Our main goal is to measure rotation periods and photometric variabilities of main-sequence stars in this large data set covering more than 500,000 stars. In particular, we investigate whether a period bimodality and/or variability decrease can also be detected in different fields of the sky. For this purpose we apply different time series analysis methods (such as Lomb-Scargle periodograms, auto-correlation functions, etc.) to the light curves to search for periodicity in a first step. The constraints in this period search are set such that as many reliable rotation periods as possible are detected. We note that these limits are comparable to previous rotation period studies using \textit{Kepler} data. In a second step, we tighten these constraints to search for a potential dearth region. We show that narrowing these constraints is crucial because in some cases spurious periods might be detected, eventually blurring the dearth region. Finally, we compare our measurements to recent studies of open clusters observed during the K2 mission.

\section{K2 data}\label{data}
The loss of the second reaction wheel ended the primary \textit{Kepler} mission, which was designed to continuously monitor more than 150,000 stars in a fixed field of view. By changing the observing strategy to monitoring different fields along the ecliptic plane (as this minimizes the torque on the spacecraft), the \textit{Kepler} telescope could be revived. This so-called K2 mission observed stars in different "Campaigns" with an observing time span of $\sim$80~days, which is comparable to the $\sim$90-day quarters of the \textit{Kepler} mission. An overview of the K2 mission concept is presented in \citet{Howell2014}. 

The K2 data for a whole campaign can be retrieved from the MAST website\footnote{http://archive.stsci.edu/pub/k2/lightcurves/tarfiles/}. In this study we analyze data from campaigns 0--18 (C0-C18)\footnote{Campaign 9 (C9) was dedicated to a gravitational microlensing study and is not considered here.}. The K2 observing strategy is to monitor each field for $\sim$70--90~days. Owing to instrumental problems (safe modes, pointing issues, etc.), the collected time series is shorter for some campaigns ($\sim$50--70~days for C10, C11, C17, C18, and $\sim$36~days for C0). Each of these campaigns contains data of up to $\sim$48,000 stars. We use light curves reduced by the PDC-MAP pipeline, which has successfully been applied to earlier \textit{Kepler} data \citep{Stumpe2012, Smith2012, Stumpe2014}. The pipeline version of the data reduction for each campaign is listed in Table~\ref{camp_table}, and the basic stellar parameters considered in this study are taken from the K2 Ecliptic Plane Input Catalog (EPIC, \citealt{Huber2016})\footnote{http://archive.stsci.edu/k2/epic/search.php}.

\section{Rotation period detection}\label{periods}
Our aim is to determine stellar rotation periods from the light curves, and to measure the photometric variability caused by the rotational modulation of active features (such as spots and faculae). In a first step, each light curve is divided by a 3rd order polynomial to account for long-term trends on the order of the time span of the data. These long-term signatures are likely remnants of an incomplete data reduction. Data points with a median absolute deviation greater than six times the median value are considered as outliers and are removed from the time series (typically up to 1\% of the number of data points). To speed up computations, we bin the data from $\sim$30-minute to 3-hours cadences. After this reduction, the photometric variability of the star is assessed by calculating the variability range $\Rvar$ from the time series. This quantity measures the difference between the 95th and 5th percentile of the sorted differential flux (see, e.g., \citealt{Basri2010,Basri2011}). For consistency, the data from each campaign are analyzed in the exact same way.

\begin{figure*}
  \centering
  \includegraphics[width=17cm]{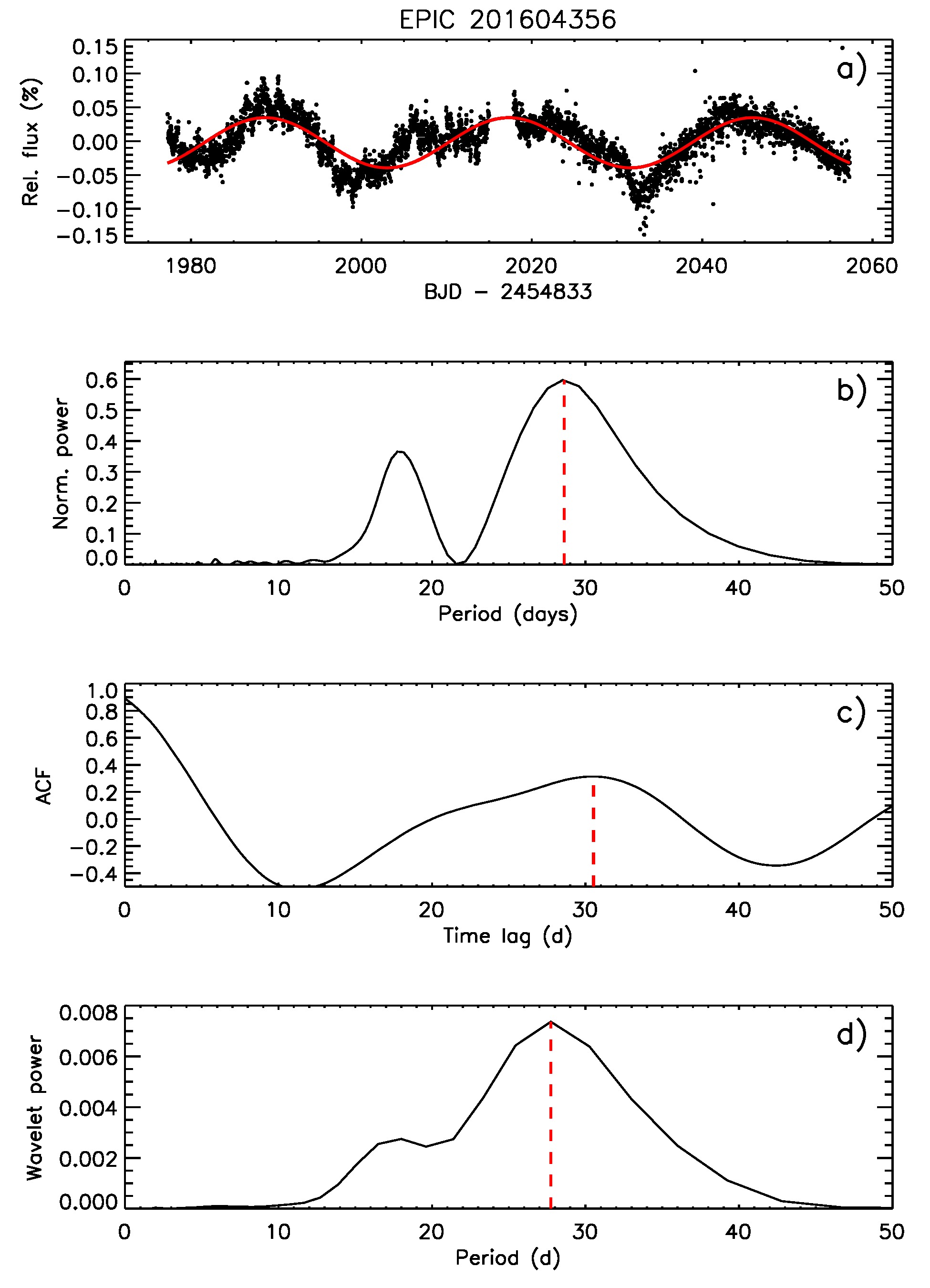}
  \caption{a) Example of the reduced light curve of the star EPIC\,201604356 observed in campaign~1. The red curve shows the best sine fit to the data. Lomb-Scargle periodogram (b), auto-correlation function (c), and integrated wavelet power spectrum (d) of the above time series. The red dashed vertical line indicates the most significant period detected by either method.}
  \label{lc}
\end{figure*}

To search for periodicity in the light curves, we apply three well-established methods of time series analysis: Lomb-Scargle periodograms, wavelet power spectra, and auto-correlation functions. The generalized Lomb-Scargle periodogram \citep{Zechmeister2009} returns peaks in frequency space. The lowest frequency is determined by the inverse of the time span of the data, and the highest frequency is given by the Nyquist frequency. The highest peak of the periodogram is associated with the frequency (or period) that minimizes the $\chi^2$ value of a sine wave fit to the data. The peak height is normalized to unity such that fitting a sine-like time series yields values close to one, whereas noisy, non-sinusoidal light curves return peak heights close to zero. The wavelet function is also based on a Fourier decomposition of the time series, tracking the periodicity over time. We use the standard 6th order Morlet wavelet, and integrate the wavelet power spectrum over time. This returns peaks on a frequency grid similar to the periodogram. The auto-correlation function (ACF) searches for self-similarity of the time series. Shifting the time series by a certain time lag (between zero and the length of the time series), and comparing it to the unshifted time series, yields the self-correlation of the time series. The time lag maximizing the correlation represents the best period in the data. An example light curve and the results of the applied methods are shown in Fig.~\ref{lc}.

Each of the methods described above returns a maximum peak at a certain period. However, these periods may differ from each other, depending on the shape of the light curve. To retrieve consistent periods among all three methods, we apply the following criteria. For the Lomb-Scargle periodogram, we require a minimum peak height of $\hpeak>0.3$. This peak height has been chosen after visual inspection of many K2 light curves from different campaigns to ensure that the signal that is picked up in the light curve is associated with rotational variability, and is not of instrumental origin. We note that this empirical limit is comparable to previous studies of rotation in \textit{Kepler} data. The period associated with this peak is our initial period guess $\Ppeak$. For short periods, $\Ppeak<10$\,d, the periods of all three methods may differ by at most one day. Intermediate periods of $10<\Ppeak<20$\,d may differ by two days, and long periods $\Ppeak>20$\,d are allowed to differ by five days. If all criteria are satisfied, we report the mean of the three periods as the rotation period $\Prot$. We note that these criteria have been set empirically by visual inspection of the light curves and the outcome of the period diagnosis methods. We require the period $\Prot$ to be longer than 1~day, and shorter than half of the observing time span. Thus, we typically derive periods in the range 1--44~days. These limits ensure that we discard very short periods, likely caused by instrumental features or stellar pulsations. Additionally, we require that at least two full rotational cycles are observed. Since we are mainly interested in rotation of main-sequence stars, we further require effective temperatures to be in the range $3250\,K < \Teff < 6250\,K$ and surface gravities $\logg > 4.2$ to exclude evolved stars. We further discard stars with anomalously high variability ranges $\Rvar>10$\%, which are likely caused by an improper data reduction. The number of stars from each campaign satisfying all constraints is given in Table~\ref{camp_table}. The "PROCVER" column contains the pipeline version that processed the data. A comparison of the results derived from different pipelines is given in appendix~\ref{app_A}.

\begin{figure}
  \resizebox{\hsize}{!}{\includegraphics{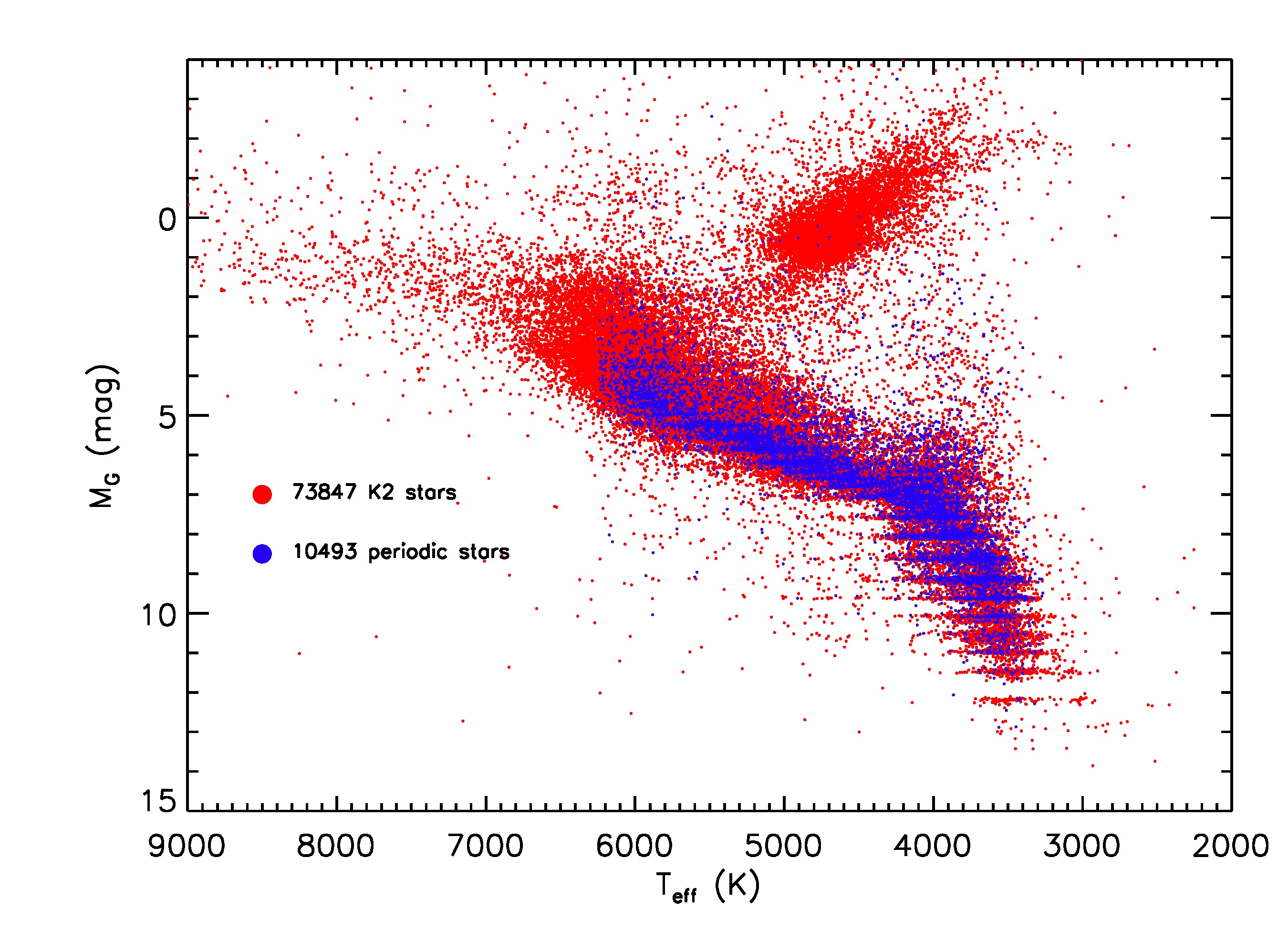}}
  \caption{Hertzsprung-Russell diagram (HRD) of all K2 stars with distances estimated by \citet{Bailer-Jones2018} and known extinctions. The blue points show stars with measured rotation periods satisfying the selection criteria from Sect.~\ref{periods}.}
  \label{HRD}
\end{figure}

Owing to the latest Gaia data release~2 (Gaia~DR2), accurate distances of a huge number of K2 stars became available \citep{Bailer-Jones2018}. To put the selected stars on a Hertzsprung-Russell diagram (HRD), we cross-matched stars from all campaigns with the Gaia DR2 catalog using a 4~arcsec search radius\footnote{https://gaia-kepler.fun/} to compute their absolute Gaia G~magnitudes. The HRD of these stars is shown in Fig.~\ref{HRD}. The red points show stars from all campaigns for which we have proper distance estimates and extinctions, and the blue points show the subsample of these stars satisfying the selection criteria used in this work. We note that the vast majority of stars selected above lies on the main sequence, where stars exhibit deep convective envelopes which are able to generate efficient magnetic dynamos. This coincidence strengthens our conclusions that the periodicity detected in these stars is indeed caused by rotational modulation.

\begin{figure}
  \resizebox{\hsize}{!}{\includegraphics{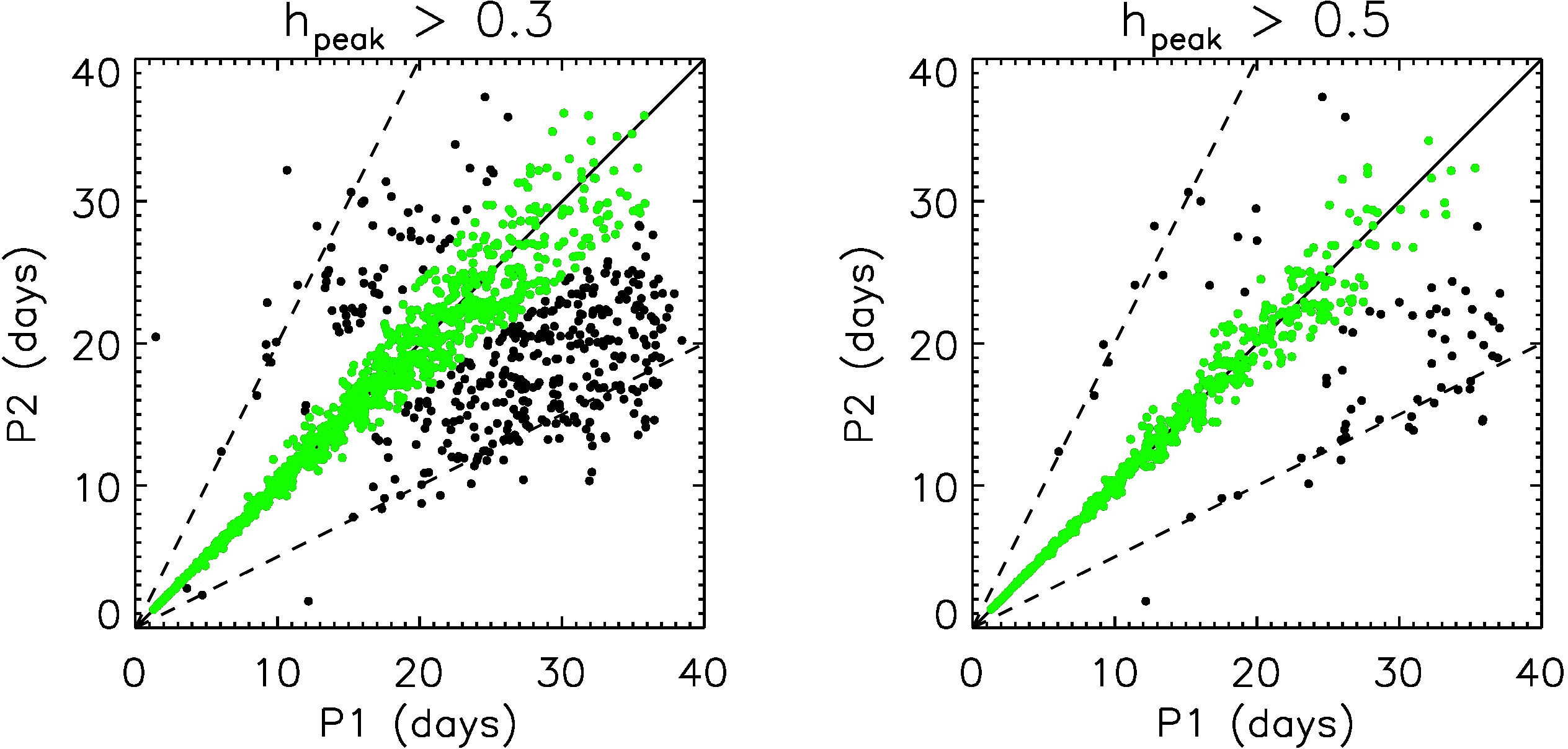}}
  \caption{Periods P1 and P2 observed in different campaigns for the same star for $\hpeak>0.3$ (left panel) and $\hpeak>0.5$ (right panel). The 1:1 ratio (solid black line), and the 1:2 and 2:1 ratios of the periods P1 and P2 (dashed black lines) are shown for guidance. Green dots show stars with a period deviation less than 20\%.}
  \label{Period12}
\end{figure}

Some parts of the K2 fields have been re-observed such that some stars have period measurements for different campaigns, with the biggest overlap between campaigns C5, C16, and C18. In our selected sample, 1861~stars have been observed more than once (1647~stars observed twice, 214~stars observed three times). These stars are now used to estimate the period uncertainties. In Fig.~\ref{Period12} we show the two periods P1 and P2 derived in different campaigns for the stars that have been observed twice. The left panel shows significant scatter for periods longer than 20~days. Periods P1 and P2 that differ by less than 20\% of the mean of P1 and P2 are shown in green, and are considered to be consistent. A deviation of 20\% is reasonable due to spot lifetimes \citep{Giles2017} and differential rotation \citep{Reinhold2013} in stars with $\Prot>20$~days. We find 1246 of the 1647~stars deviating by less than 20\%, which amounts to a quite large fraction of 75.7\% of stars. The large deviations for some stars can be explained because the periods P1 and P2 were derived from campaigns with different observing time span. One can see that P1 is, on average, up to 50\% longer than P2. This accounts for the fact that P1 refers to an earlier campaign than P2, and that the average observing time span was longer for earlier campaigns (see Sect.~\ref{data}). Thus, longer periods are picked up more easily for longer observing windows. However, the period reliability can be increased by setting tighter constraints in the period search. The right panel in Fig.~\ref{Period12} shows the same as the left panel now only for stars with $\hpeak>0.5$. This quite high limit significantly reduces the scatter for periods above 20~days, and heavily shrinks the sample size. For $\hpeak>0.5$ we only find 71 out of 733~stars deviating by more than 20\%, which is equal to 9.7\%, i.e. for more than 90\% we find consistent rotation periods.

\begin{figure}
  \resizebox{\hsize}{!}{\includegraphics{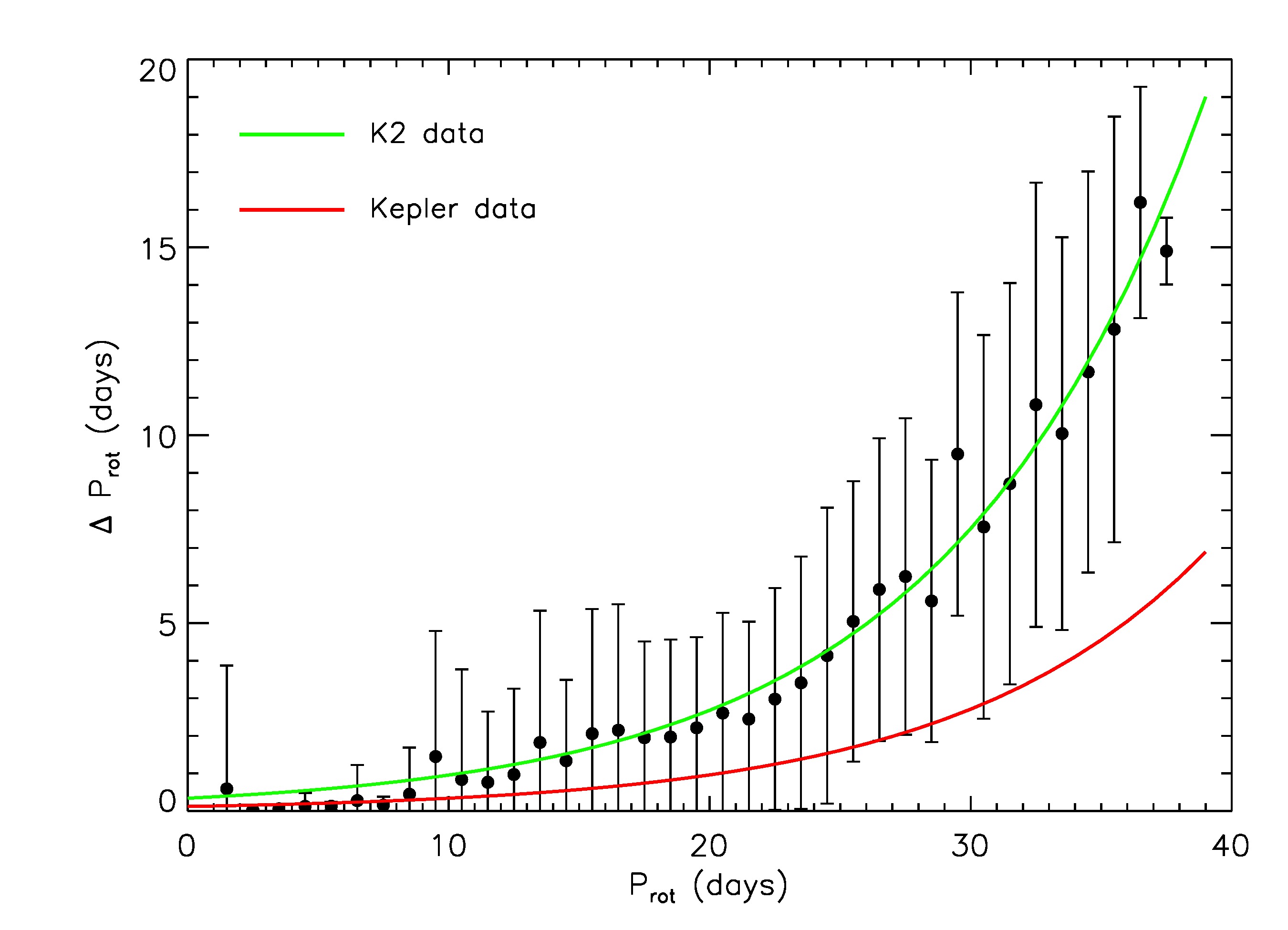}}
  \caption{Estimating the rotation period uncertainty for the case $\hpeak>0.3$. The measured rotation period $\Prot$ is given on the x-axis. On the y-axis, the mean and standard deviation of the absolute difference of the periods P1 and P2 is given within period bins of one day. The green curve shows a fit to the data (using equal weights), and the red curve shows a similar fit estimated from \textit{Kepler} data (see Fig.~\ref{uncertainties_kepler} in the appendix).}
  \label{delta_Period12}
\end{figure}

To estimate rotation period uncertainties, we calculate the absolute difference of the periods P1 and P2, over the full period range in period bins of one day, and compute the mean and standard deviation within each period bin. This is shown in Fig.~\ref{delta_Period12}. The green curve shows an exponential fit to the data using equal weights (the error bars are only shown to visualize the spread in each period bin). The red curve shows the same function comparing the period derived from a single \textit{Kepler} quarter to the result obtained from the full light curve (see appendix~\ref{method_Kepler}). The period uncertainty of the K2 stars is much larger because two individual campaigns (of roughly equal length) have been compared with each other. The periods of the \textit{Kepler} stars, however, are more reliable since the time base of observations is much longer. The red curve is provided here to show that our methods are reasonable and yield reliable results. The exponential fit to the K2 periods (green curve) can be exploited to derive period uncertainties $\Delta\Prot$ at the rotation periods $\Prot$. The red curve should be considered as a lower limit. The measured rotation periods, uncertainties, and other fundamental stellar parameters are given in Table~\ref{par_table} for the first ten stars in our sample. Note that the full table contains \nperiods~periods because several stars have been observed more than once.

\begin{table*}
  \centering
  \begin{tabular}{ccccccc}
\hline\hline
Campaign & PROCVER & No. of stars & No. of stars $h_{peak}>0.3$ & \% of sel. stars 
& No. of stars $h_{peak}>0.5$ & \% of sel. stars \\
\hline
0 & 9.3.97 & 1886  & 308  & 16.3 & 162  & 8.6 \\ 
1 & 9.3.58 & 14074 & 2871 & 20.4 & 1146 & 8.1 \\ 
2 & 9.3.85 & 3022  & 591  & 19.6 & 302  & 10.0 \\ 
3 & 9.3.15 & 8831  & 1648 & 18.7 & 732  & 8.3 \\  
4 & 9.3.19 & 7776  & 1708 & 22.0 & 911  & 11.7 \\  
5 & 9.3.31 & 13118 & 3510 & 26.8 & 1848 & 14.1 \\ 
6 & 9.3.40 & 16672 & 2470 & 14.8 & 1013 & 6.1 \\ 
7 & 9.3.45 & 3894  & 629  & 16.2 & 315  & 8.1 \\ 
8 & 9.3.46 & 15093 & 2612 & 17.3 & 1051 & 7.0 \\ 
9 & - & - & - & - & - & - \\
10 & 9.3.60 & 17459 & 2045 & 11.7 & 749  & 4.3 \\
11 & 9.3.67 & 2945  & 536  & 18.2 & 277  & 9.4 \\
12 & 9.3.70 & 19643 & 2182 & 11.1 & 708  & 3.6 \\
13 & 9.3.72 & 8892  & 1479 & 16.6 & 689  & 7.7 \\
14 & 9.3.75 & 11636 & 1575 & 13.5 & 722  & 6.2 \\
15 & 9.3.84 & 10800 & 1552 & 14.4 & 702  & 6.5 \\
16 & 9.3.87 & 13219 & 2439 & 18.5 & 1212 & 9.2 \\
17 & 9.3.90 & 18991 & 2052 & 10.8 & 711  & 3.7 \\
18 & 9.3.93 & 11011 & 2180 & 19.8 & 1095 & 9.9 \\
\hline
\end{tabular}

  \caption{Selected number of stars from each campaign satisfying all constraints from Sect.\ref{periods}. The third column contains the number of stars satisfying the temperature, log\,g, and variability cuts. The "PROCVER" column shows the pipeline version that the data were processed with.}
  \label{camp_table}
 \end{table*}
 
\begin{table*}
  \centering
  \begin{tabular}{cccccccccc}
\hline\hline
EPIC & Campaign & $T_{\rm eff}$ & $\log\,g$ & $P_{\rm rot}$ & $\Delta P_{\rm rot}$ & $h_{\rm peak}$ & $R_{\rm var}$ & Kp & $\rm M_G$ \\
 &  & (K) & (dex) & (days) & (days) &  & (\%) & (mag) & (mag) \\
\hline
202059193 & 0 & 3832 & 4.88 & 15.40 & 1.67 & 0.76 & 0.28 & 12.40 & 8.08 \\
202059198 & 0 & 4163 & 4.79 & 16.51 & 1.87 & 0.47 & 0.37 & 11.20 & - \\
202059199 & 0 & 3808 & 4.95 & 16.89 & 1.95 & 0.34 & 0.26 & 12.20 & - \\
202059204 & 0 & 3718 & 4.95 & 7.93 & 0.77 & 0.66 & 2.82 & 11.30 & - \\
202059207 & 0 & 3858 & 4.94 & 15.45 & 1.68 & 0.61 & 0.06 & 12.20 & - \\
202059210 & 0 & 5038 & 4.61 & 15.57 & 1.70 & 0.85 & 1.48 & 11.80 & - \\
202059224 & 0 & 3389 & 5.10 & 13.28 & 1.34 & 0.46 & 0.07 & 11.60 & - \\
202059229 & 0 & 4161 & 4.78 & 4.93 & 0.57 & 0.44 & 2.39 & 10.50 & 8.06 \\
202059231 & 0 & 3856 & 4.88 & 17.95 & 2.17 & 0.73 & 3.55 & 11.70 & - \\
202059586 & 0 & 5915 & 4.37 & 4.32 & 0.53 & 0.62 & 1.18 & 14.60 & - \\
\hline
\end{tabular}

  \caption{Measured rotation periods and stellar fundamental parameters from the EPIC for 10~stars in our sample. The last two columns contain the apparent \textit{Kepler} magnitude Kp and the absolute Gaia magnitude $\rm M_G$. For the stars with missing $\rm M_G$ values either no distances and/or extinctions were available. The full table can be obtained from the CDS.}
  \label{par_table}
\end{table*}

\section{Results}
As described in the previous section, some stars have been observed during different campaigns. For all stars observed more than once, in the following we only consider the mean value of the rotation periods, period uncertainties, peak heights, and photometric variabilities measured in different campaigns. Stars with period deviations $>20\%$ among different observing campaigns have been excluded from the sample. In total, \nstars~stars satisfy all selection criteria. 

\subsection{Rotation period distribution}
\begin{figure*}
  \centering
  \includegraphics[width=17cm]{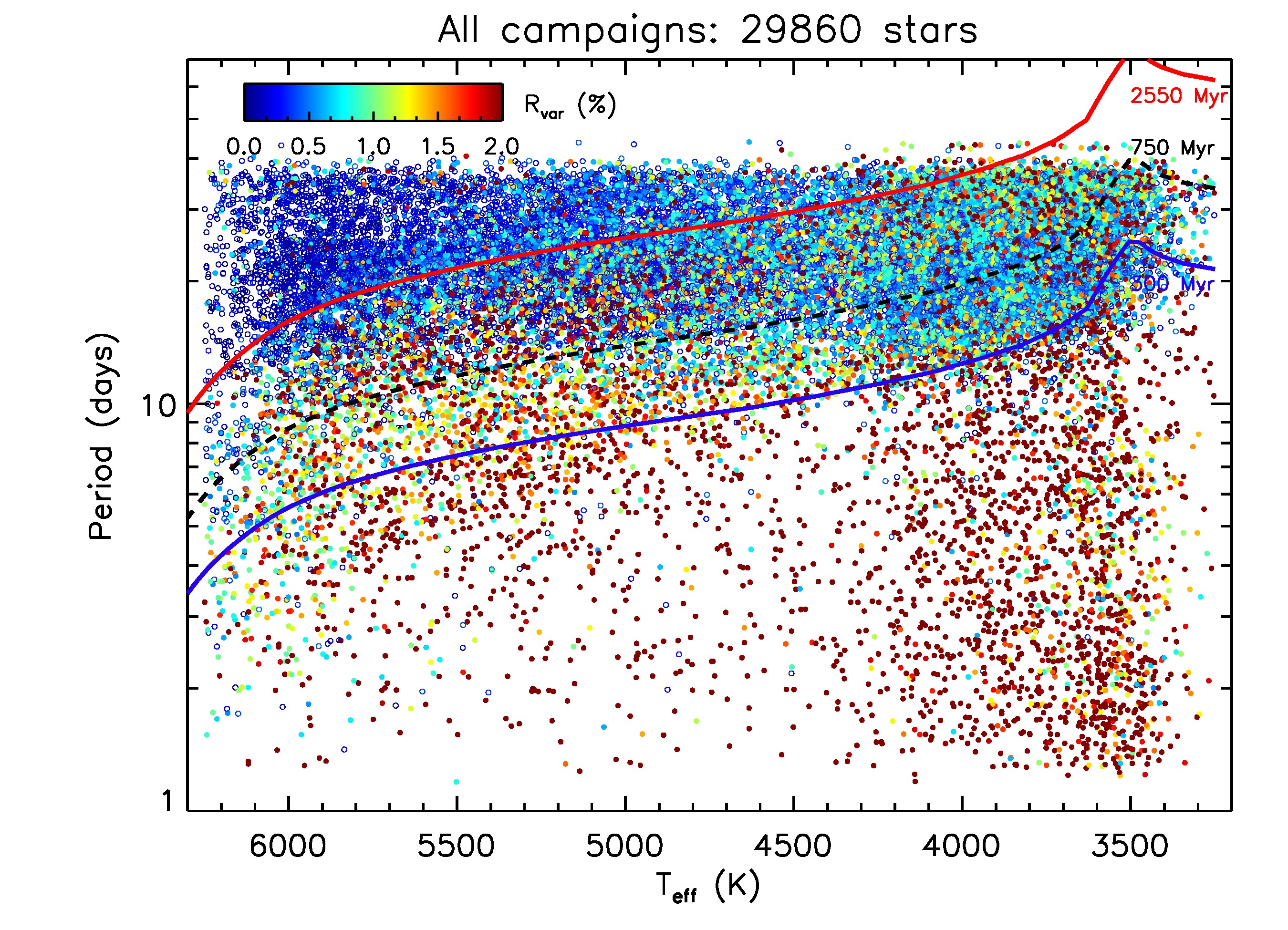}
  \caption{Rotation periods versus effective temperature of all stars with peak heights $\hpeak>0.3$. The data are color-coded with the variability range $\Rvar$. Stars with $\Rvar<0.5$\% are shown as open circles to increase visibility. The blue, black, and red line show isochrones of 300\,Myr, 750\,Myr, and 2550\,Myr, respectively, using gyrochronology relations from \citet{Barnes2010}.}
  \label{Teff_Prot}
\end{figure*}

We now turn to the rotation period distribution. As mentioned above, stars with measured rotation periods are thought to hold magnetic dynamos. Owing to charged particles following the magnetic field lines (i.e. the stellar wind, see e.g. \citealt{Mestel1968,Mestel1968b}), stars spin down with age \citep{Skumanich1972}. The efficiency of this magnetic braking depends on stellar mass \citep{Barnes2003, Meibom2011}. In Fig.~\ref{Teff_Prot} we show the measured rotation periods as a function of effective temperature (a proxy for stellar mass). The data are color-coded by the measured variability range $\Rvar$ (we return to this quantity in Sect.~\ref{variability}). The colored lines show isochrones predicted from gyrochronology relations for three different ages (Eq.~9 in \citealt{Barnes2010}).

We find that the rotation periods increase towards cooler stars. The data show good agreement with the shape of the isochrones, and the 300\,Myr isochrone represents a lower age limit for the majority of stars. For stars cooler than $\sim$4000\,K, however, a large fraction of fast rotators with high variability ranges are present. This observation is consistent with the decreased efficiency of magnetic braking for M~dwarfs (see e.g. \citealt{Reiners2012} and the discussion in Sect.~\ref{discussion}). On the other side of the period distribution, we are limited to periods up to $\sim$40~days by the time span of the data, although we expect stars to rotate even slower, especially the M~dwarfs for which periods longer than 100~days have been detected \citep{Irwin2011, Newton2016}.

The majority of stars below the 300\,Myr isochrone show high photometric variabilities, whereas stars above the 2550\,Myr isochrone exhibiting small variabilities. This general trend is not surprising: younger stars rotate faster and exhibit higher magnetic activity than older slowly-rotating stars \citep{Noyes1984}. Consequently, fast rotators exhibit stronger magnetic fields able to generate larger, long-lived surface features (such as dark spots and bright faculae), causing higher photometric variability in the light curves.

\begin{figure*}
  \centering
  \includegraphics[width=17cm]{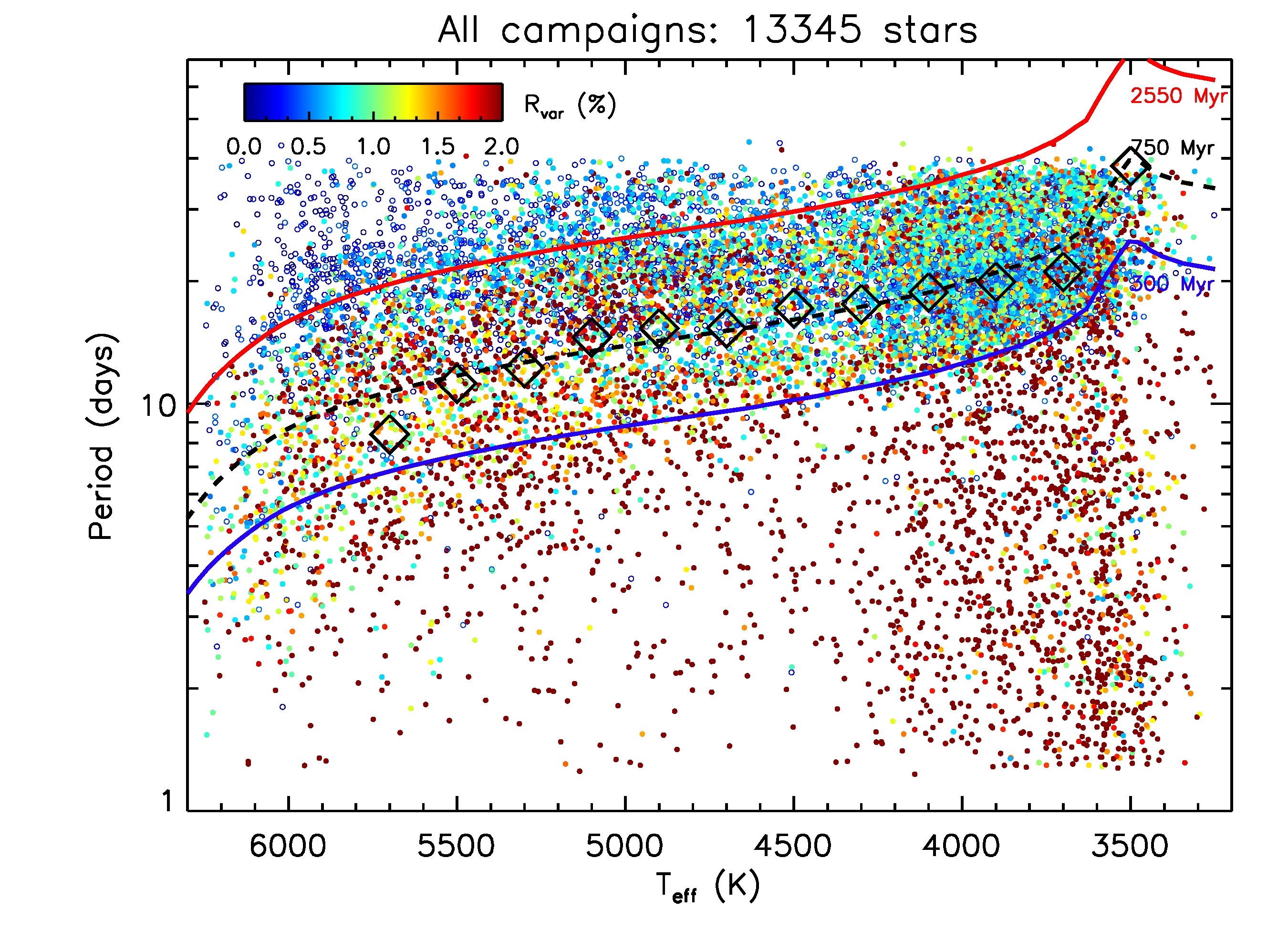}
  \caption{Rotation periods versus effective temperature of all stars with peak heights $\hpeak>0.5$. The color-coding and isochrones are the same as in Fig.~\ref{Teff_Prot}. The black diamonds show periods where a local minimum in the variability range was detected (see Fig.~\ref{Prot_range} and Sect.~\ref{variability}).}
  \label{Teff_Prot2}
\end{figure*}

To investigate the potential presence of a dearth region at intermediate rotation periods, we now tighten the constraints in the period search to a peak height limit of $\hpeak>0.5$, while leaving all other constraints unchanged. Increasing the peak height limit to $\hpeak>0.5$ is crucial because the reliability of these periods is much better, with large deviations only occurring in less than 10\% of the cases (see Fig.~\ref{Period12}). In Fig.~\ref{Teff_Prot2} we show the rotation period distribution for all stars satisfying $\hpeak>0.5$. This criterion reduces the sample to \nstarstwo~stars, with the most dramatic decrease of stars above the 2550\,Myr isochrone exhibiting small variabilities. The dearth region is now more visible as compared to Fig.~\ref{Teff_Prot}, which we attribute to the better period accuracy. For all stars satisfying these much stricter constraints, we now focus on their distribution of photometric variabilities. 

\subsection{Variability decrease around 750\,Myr}\label{variability}
Taking a closer look at the variability distribution in Fig.~\ref{Teff_Prot} and Fig.~\ref{Teff_Prot2}, it shows that the variability range $\Rvar$ does not decrease monotonically towards longer periods but shows a local minimum along the 750\,Myr isochrone. For a fixed temperature, the variability decreases with increasing period towards the 750\,Myr isochrone, then slightly increases again, and eventually decreases towards long periods. This behavior is better visible in Fig.~\ref{Prot_range}. We show the logarithm of the variability range $\Rvar$ versus rotation period $\Prot$ for stars in 200\,K temperature bins spanning the range 3400--5800\,K. The general decrease of variability with increasing rotation period is clearly visible for all temperature bins, although the data show large scatter. Therefore, we compute the median period for period bins of 1~day and the standard error (i.e. the standard deviation divided by the square root of the number of data points in that bin), shown as blue dots and error bars, and fit them with a spline function (solid blue line). At certain periods, this function shows a local minimum, indicated by the dashed blue line. As temperature increases (left to right and top to bottom in Fig.~\ref{Prot_range}), this local minimum is shifted to shorter periods. For temperatures in the range 4800--5600\,K the minimum is less pronounced, and the variability curve rather shows a plateau shape. For these temperature bins, we indicate the inflection point. The periods associated with these local minima of $\Rvar$ are shown as open black diamonds in Fig.~\ref{Teff_Prot2}. These periods and associated mean temperatures can be turned into gyrochronology ages using the relations from \citet{Barnes2010}. The derived ages range from 460--880\,Myr, with a median and standard deviation of $750\pm140$\,Myr. Note that a similar dependence between the photometric variability and the rotation period was found for \textit{Kepler} data (see Figs.~7-9 in \citealt{Basri2018}).


The 750\,Myr isochrone shows good agreement with the region of decreased variability in Fig.~\ref{Teff_Prot} and Fig.~\ref{Teff_Prot2}. However, the periods at the local variability minima in Fig.~\ref{Prot_range}, that were used to derive the age of 750\,Myr, are accompanied by large uncertainties because the minima are rather broad. Table~\ref{Prot_uncertainties} shows the rotation periods predicted by the gyrochronology relations from \citet{Barnes2010} for an isochrone of 750\,Myr at the mean values of the chosen temperature bins. The period uncertainties $\Delta P_{\rm rot}$ were derived by computing the exponential function from Fig.~\ref{delta_Period12} at the predicted gyrochronology periods. We note that the large period uncertainty for the lowest temperature bin in Table~\ref{Prot_uncertainties} should be interpreted as an alias detection, i.e. the period P2 roughly equals half the period P1 (see Fig.~\ref{Period12}). Moreover, the data at these long periods are sparse, and close to our upper period detection limit. The period uncertainties in Table~\ref{Prot_uncertainties} are used in the following section to see whether the period at minimum variability coincides with a potential bimodality in the period distribution. 
\begin{table*}
  \centering
  \begin{tabular}{ccc}
\hline\hline
$T_{\rm eff}$ & $P_{\rm rot}$ & $\Delta P_{\rm rot}$ \\
\hline
3500 & 39.96 & 23.55 \\
3700 & 24.65 & 3.60 \\
3900 & 20.98 & 2.30 \\
4100 & 18.75 & 1.75 \\
4300 & 17.30 & 1.46 \\
4500 & 16.09 & 1.26 \\
4700 & 15.10 & 1.12 \\
4900 & 14.27 & 1.01 \\
5100 & 13.47 & 0.91 \\
5300 & 12.62 & 0.82 \\
5500 & 11.72 & 0.74 \\
5700 & 10.71 & 0.65 \\
\hline
\end{tabular}

  \caption{Predicted periods $\Prot$ and uncertainties $\Delta P_{\rm rot}$ for an isochrone of 750\,Myr at different effective temperatures.}
  \label{Prot_uncertainties}
 \end{table*}

\subsection{Period bimodality}
\begin{figure*}
  \centering
  \begin{minipage}[b]{0.33\textwidth}
  \includegraphics[width=\textwidth]{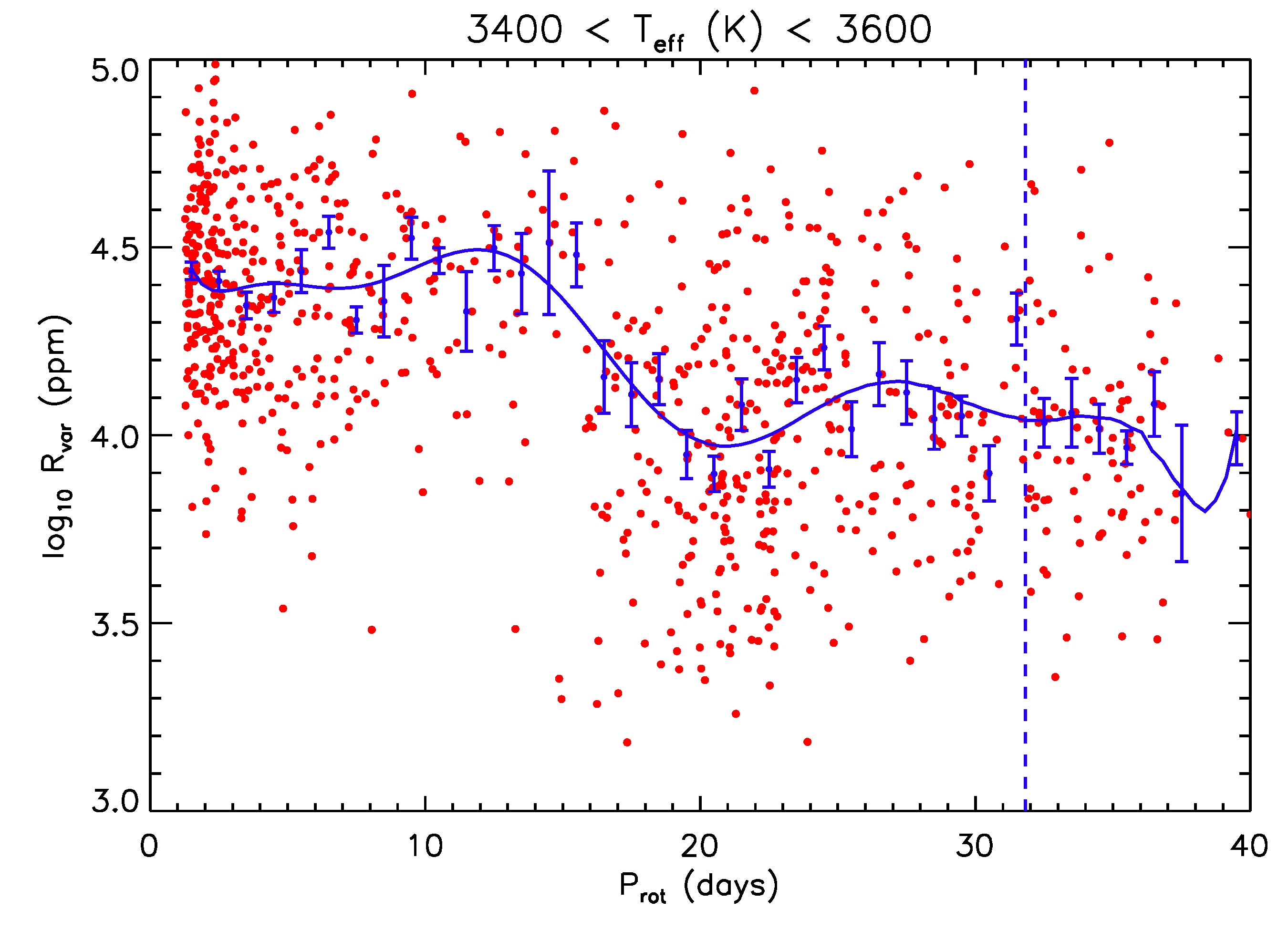}
  \end{minipage}
  \hfill
  \begin{minipage}[b]{0.33\textwidth}
  \includegraphics[width=\textwidth]{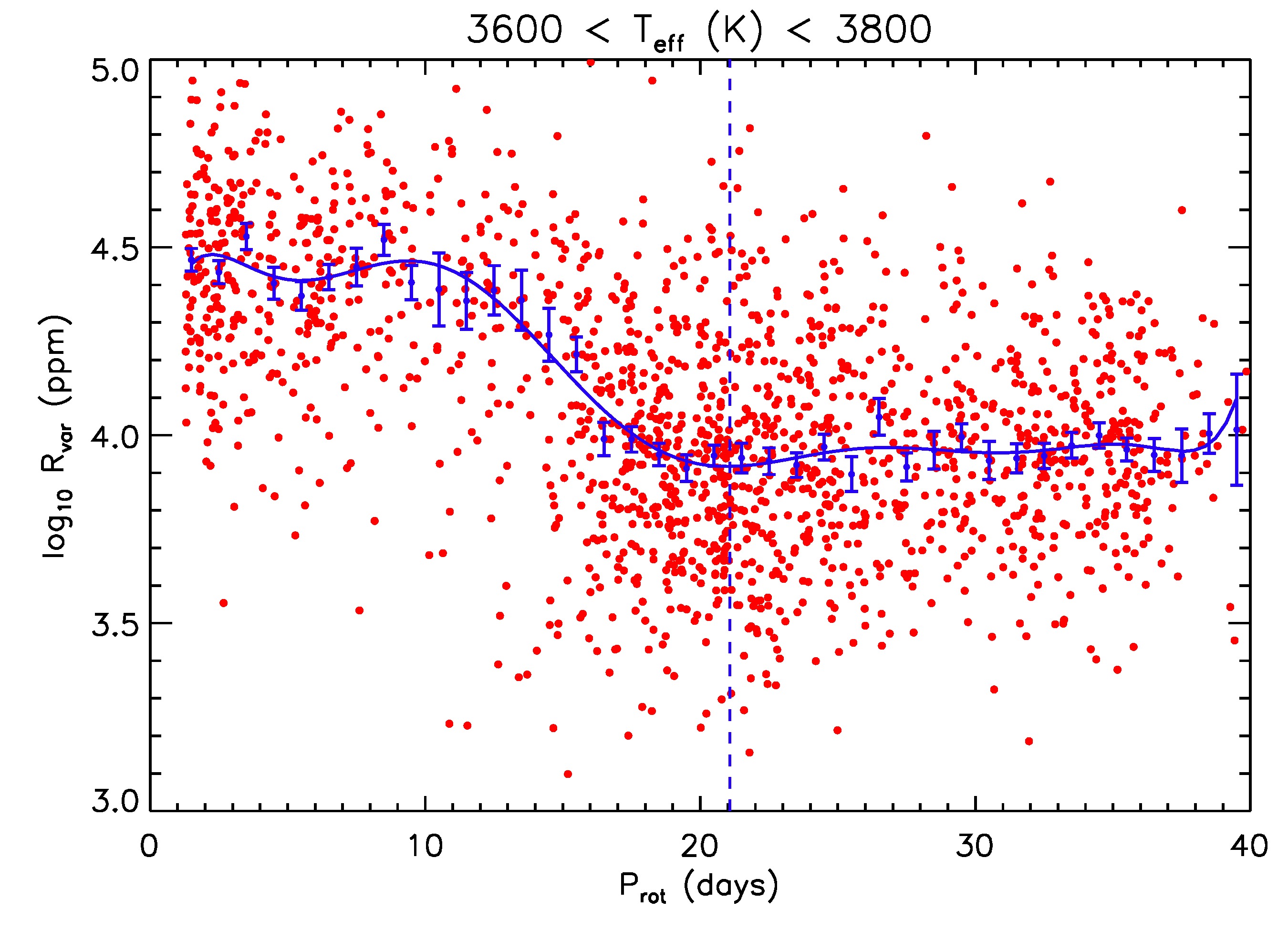}
  \end{minipage}
  \hfill
  \begin{minipage}[b]{0.33\textwidth}
  \includegraphics[width=\textwidth]{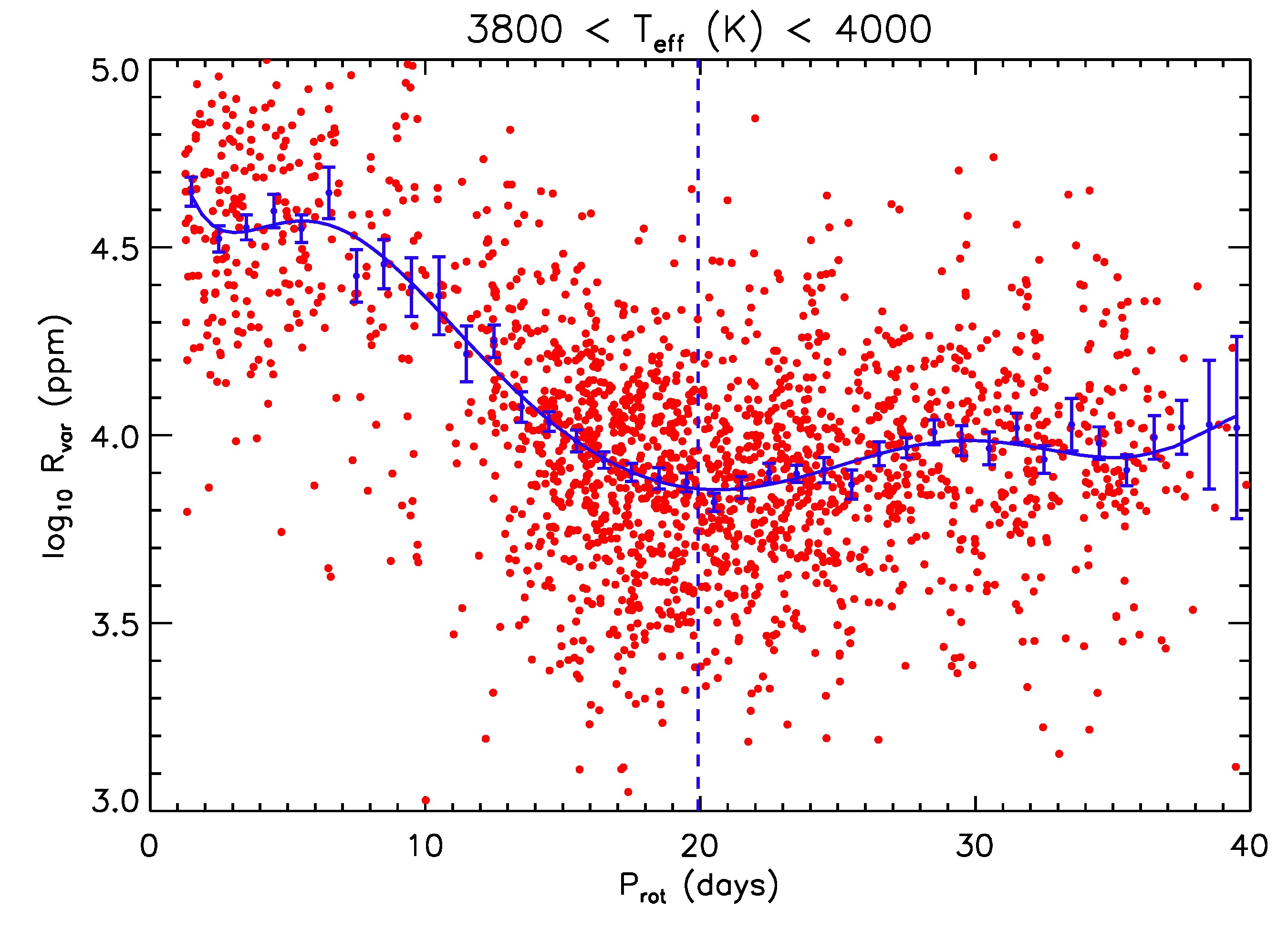}
  \end{minipage}
  \begin{minipage}[b]{0.33\textwidth}
  \includegraphics[width=\textwidth]{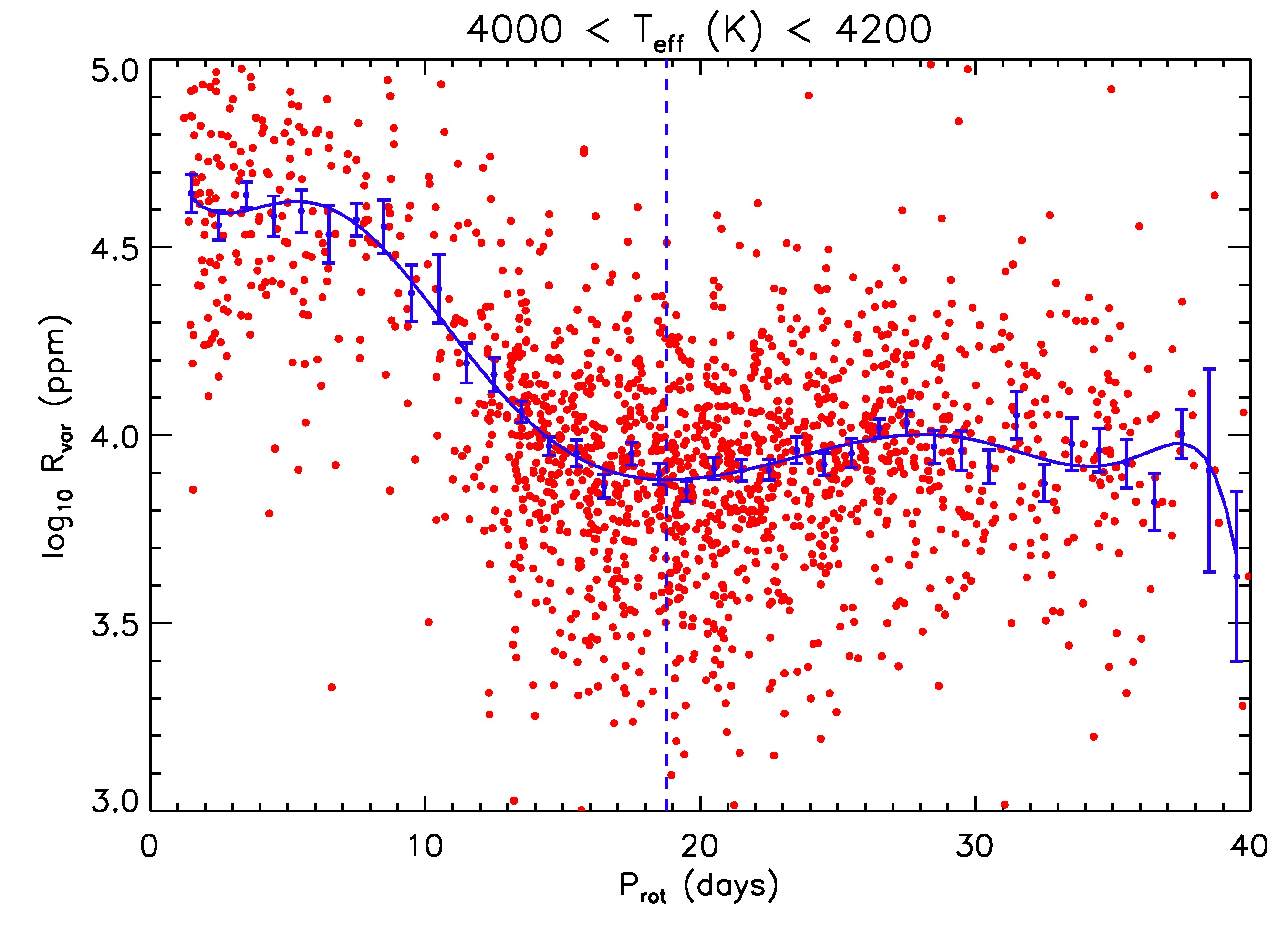}
  \end{minipage}
  \hfill
  \begin{minipage}[b]{0.33\textwidth}
  \includegraphics[width=\textwidth]{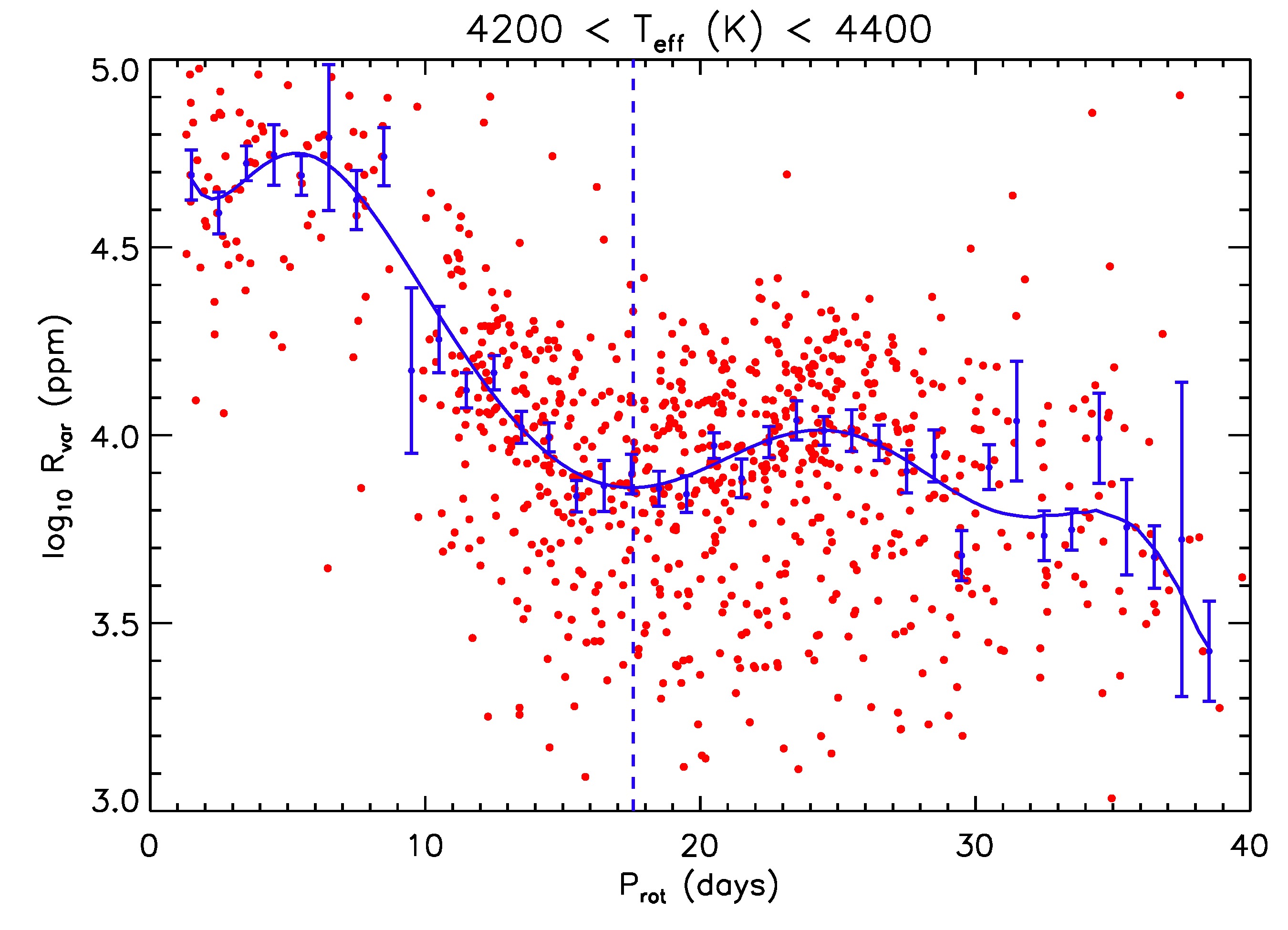}
  \end{minipage}
  \hfill
  \begin{minipage}[b]{0.33\textwidth}
  \includegraphics[width=\textwidth]{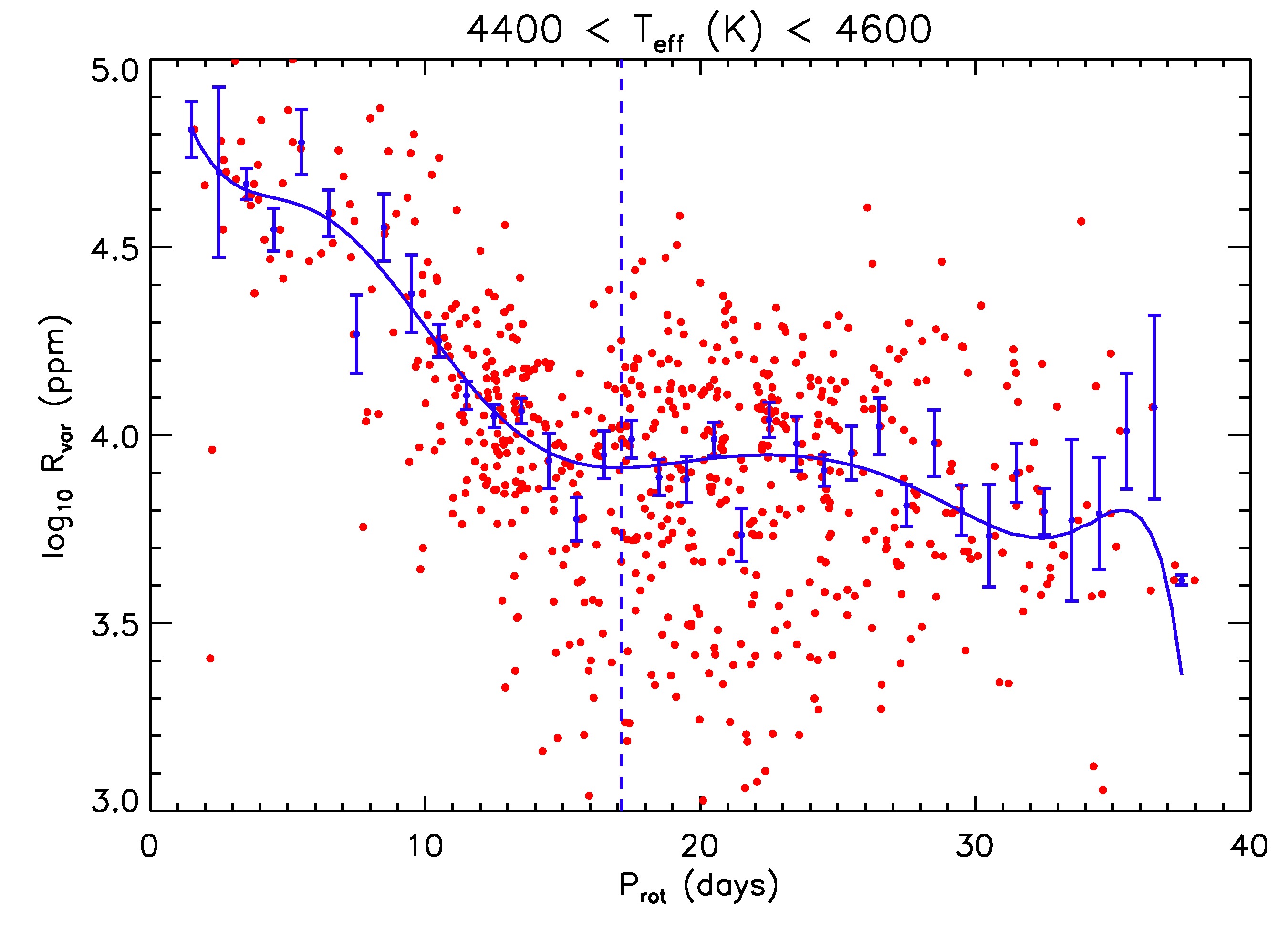}
  \end{minipage}
  \begin{minipage}[b]{0.33\textwidth}
  \includegraphics[width=\textwidth]{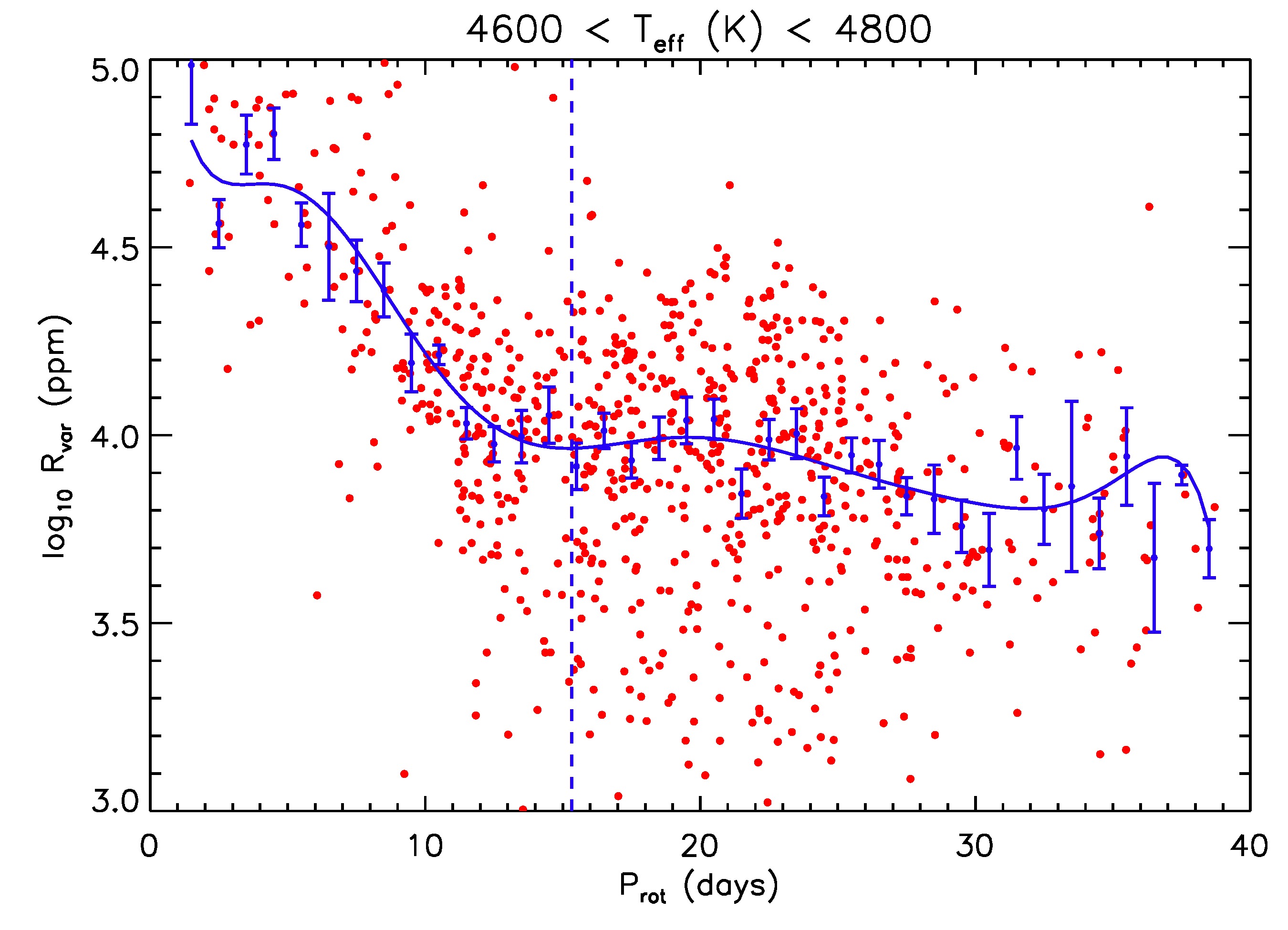}
  \end{minipage}
  \hfill
  \begin{minipage}[b]{0.33\textwidth}
  \includegraphics[width=\textwidth]{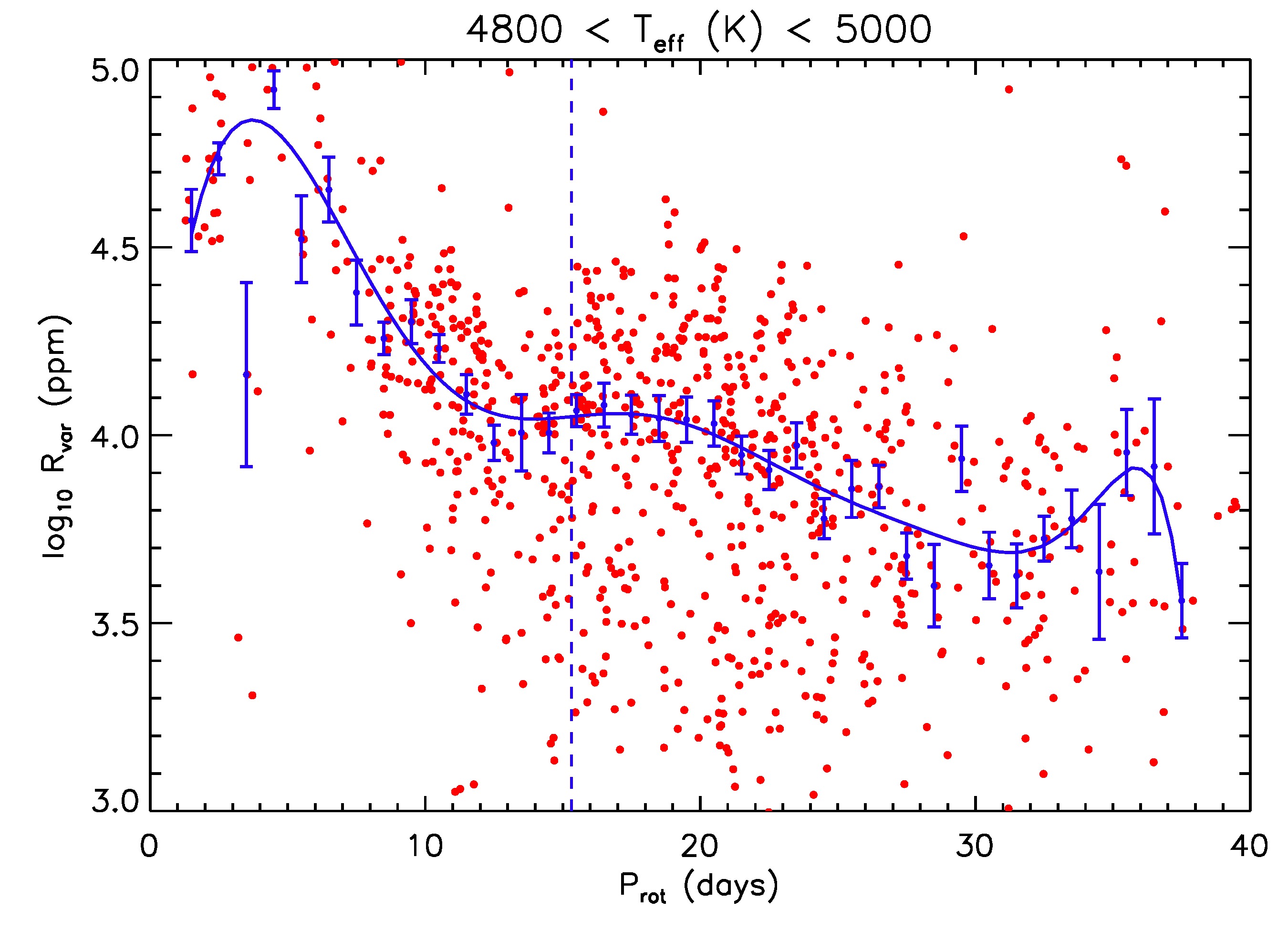}
  \end{minipage}
  \hfill
  \begin{minipage}[b]{0.33\textwidth}
  \includegraphics[width=\textwidth]{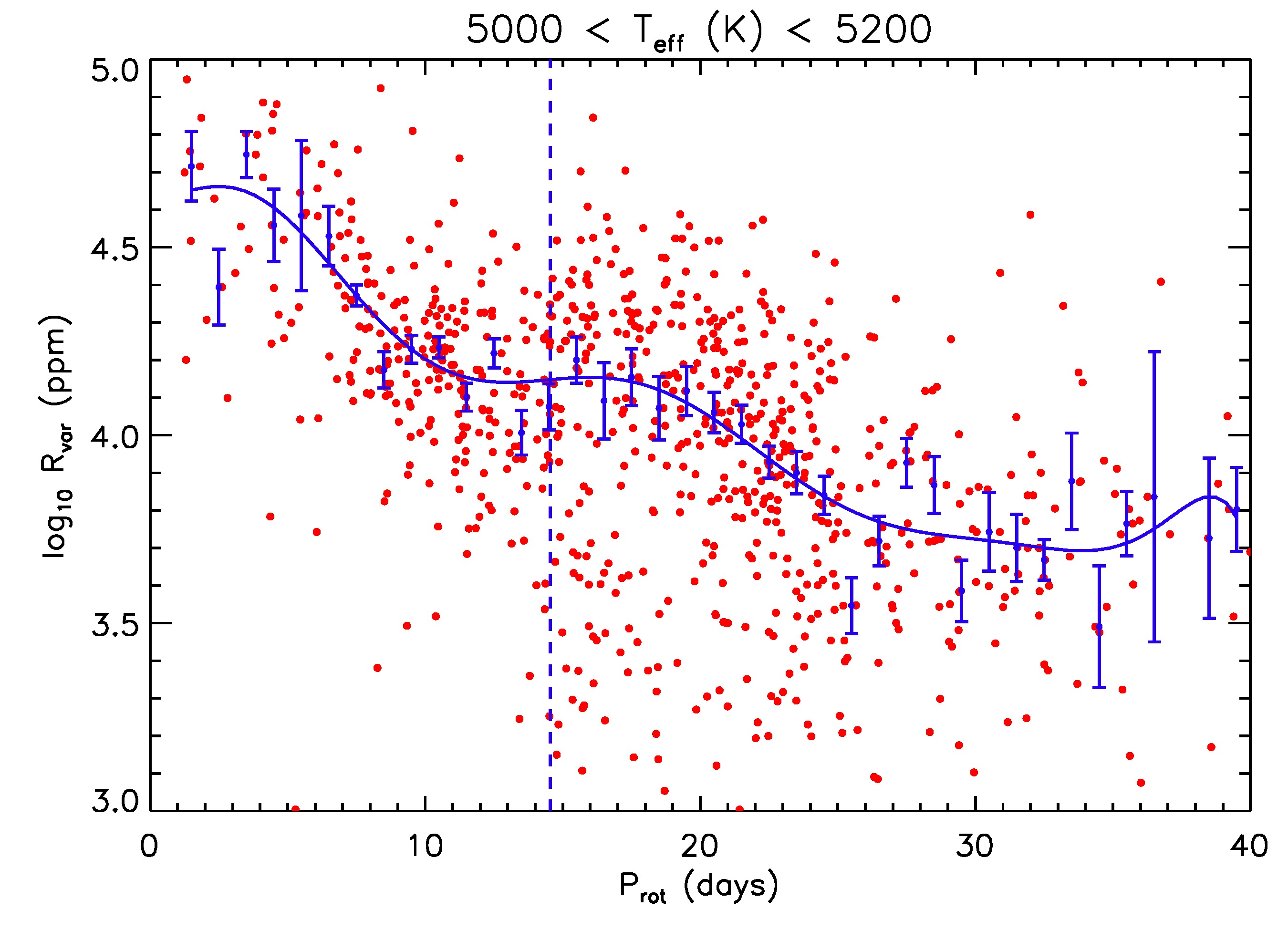}
  \end{minipage}
  \begin{minipage}[b]{0.33\textwidth}
  \includegraphics[width=\textwidth]{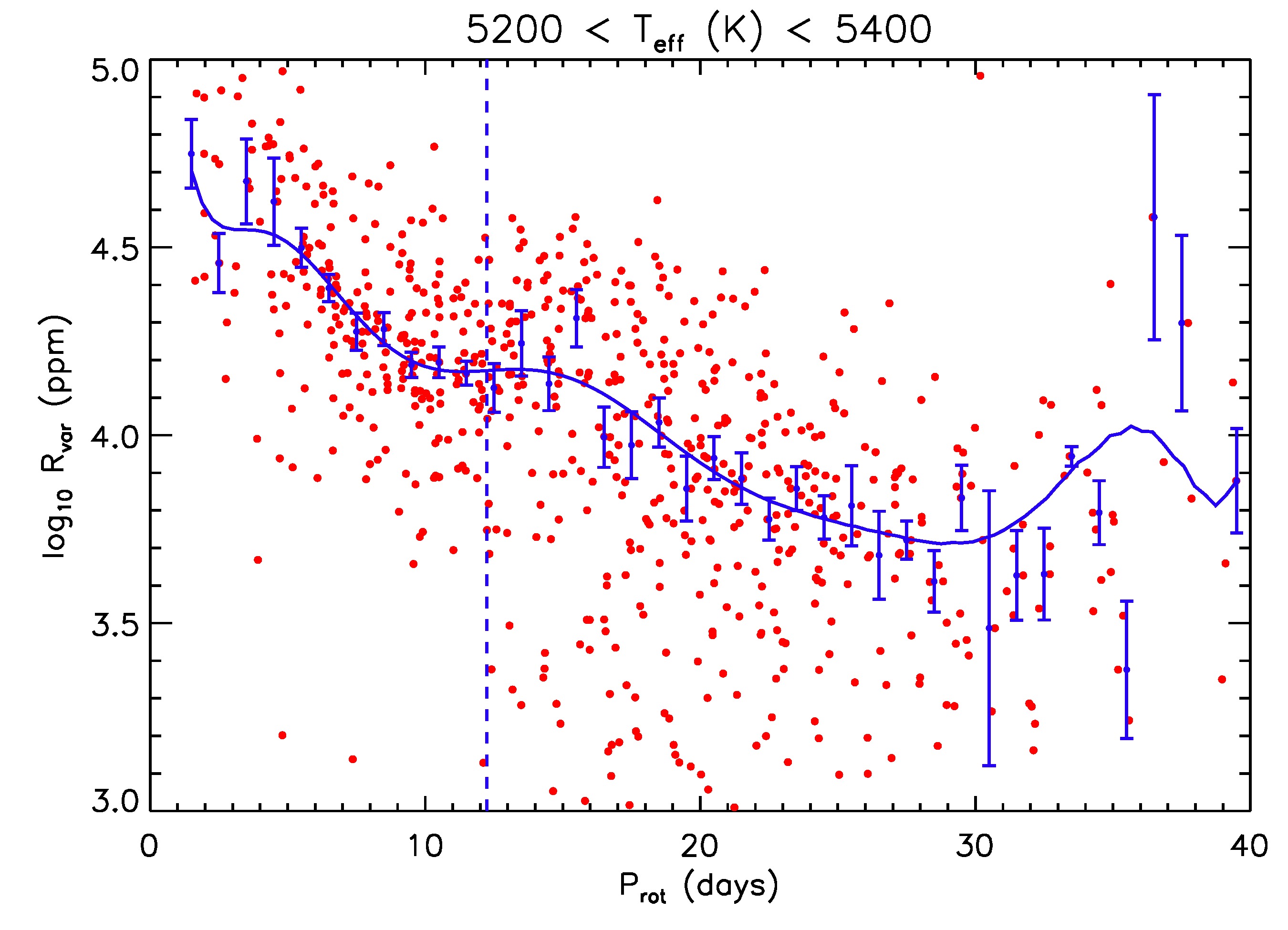}
  \end{minipage}
  \hfill
  \begin{minipage}[b]{0.33\textwidth}
  \includegraphics[width=\textwidth]{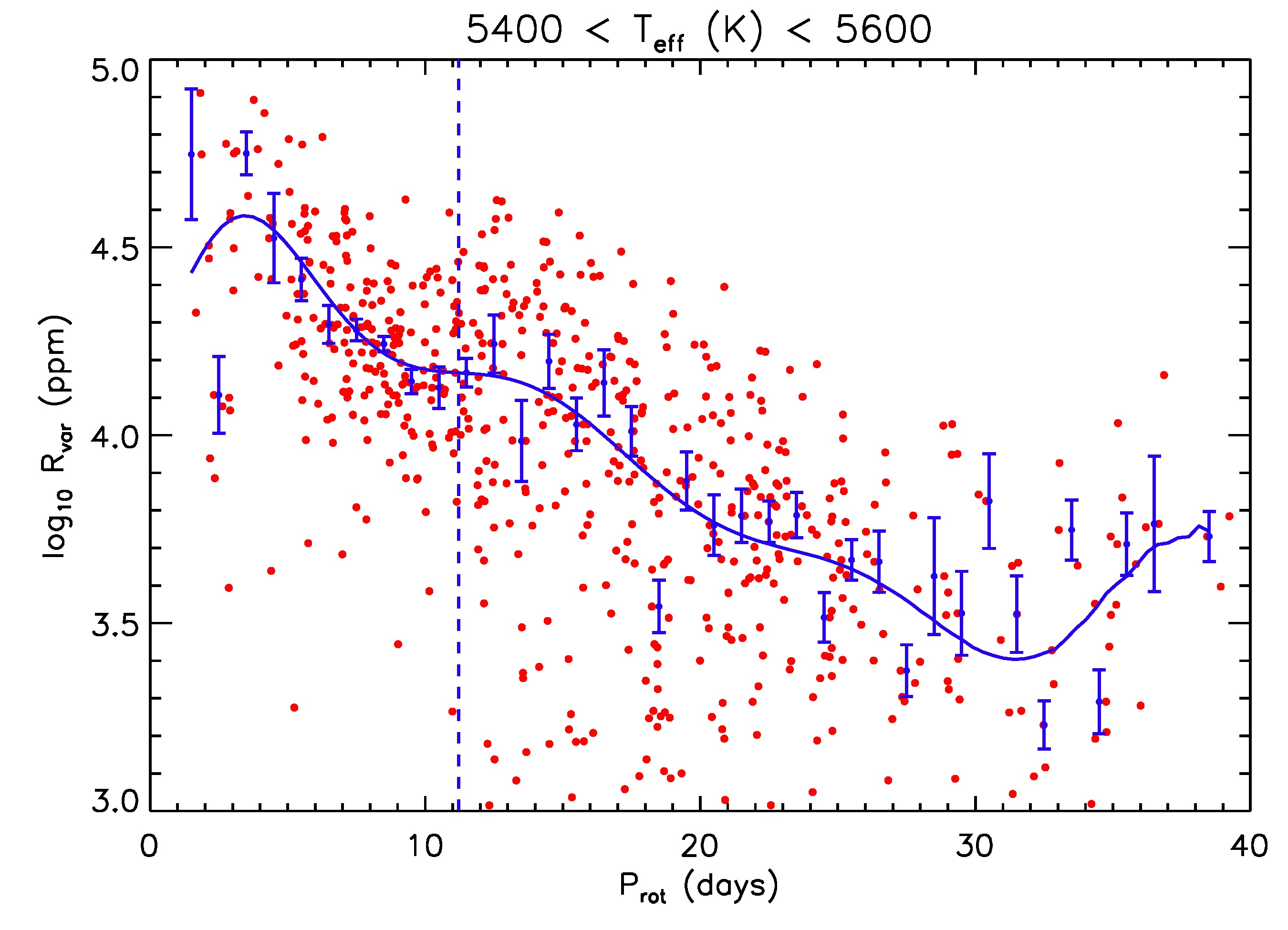}
  \end{minipage}
  \hfill
  \begin{minipage}[b]{0.33\textwidth}
  \includegraphics[width=\textwidth]{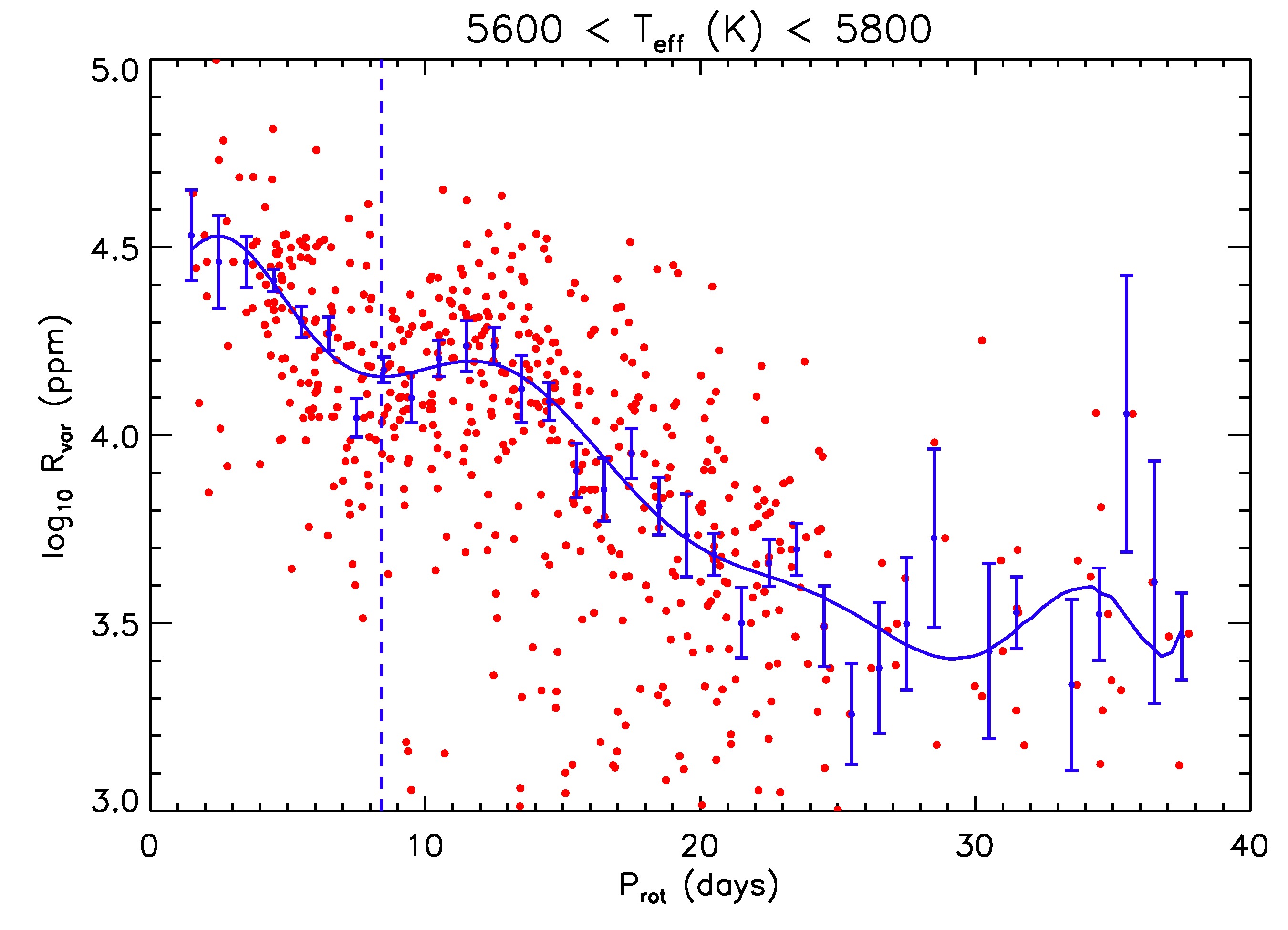}
  \end{minipage}
  \caption{Variability range $\Rvar$ versus rotation period $\Prot$ for different temperature bins (red dots). The blue dots show the median variability range for period bins of one day, and the error bars indicate the standard error therein. The blue solid line shows a spline fit to the median values. The blue dashed vertical lines indicate local minima of the variability range.}
  \label{Prot_range}
\end{figure*}

We now return to the idea proposed by \citet{Reinhold2019} to explain the previously observed lack of stars with intermediate rotation periods in the \textit{Kepler} field. These authors found that the dearth region, where significantly less rotation periods were detected, shows good agreement with an 800\,Myr isochrone. Additionally, the variability along this isochrone shows a local minimum. This observation is consistent with the behavior of the K2 stars along the 750\,Myr isochrone. To test whether the variability decrease in our sample is also accompanied by a lack of detected rotation periods, we analyze the rotation period distribution for the same temperature bins as in Fig.~\ref{Prot_range}. Following the 750\,Myr isochrone from low to high temperatures, the period distribution is expected to show some bimodality, with the dip of the distribution moving from $\sim$30~days for cooler stars to shorter periods down to $\sim$10~days for hotter stars. 

\begin{figure*}[t]
  \centering
  \begin{minipage}[b]{0.33\textwidth}
  \includegraphics[width=\textwidth]{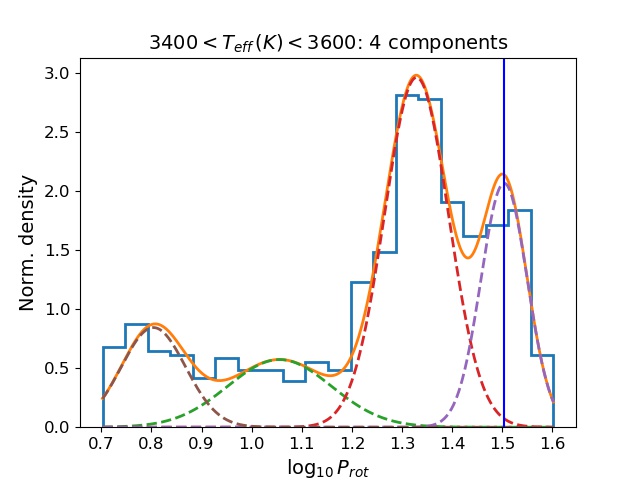}
  \end{minipage}
  \hfill
  \begin{minipage}[b]{0.33\textwidth}
  \includegraphics[width=\textwidth]{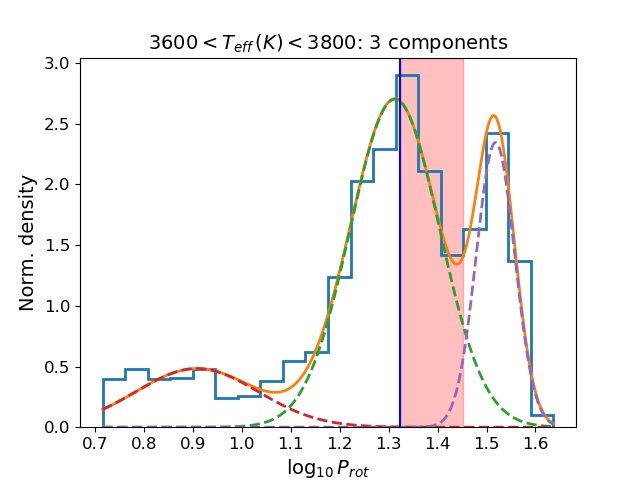}
  \end{minipage}
  \hfill
  \begin{minipage}[b]{0.33\textwidth}
  \includegraphics[width=\textwidth]{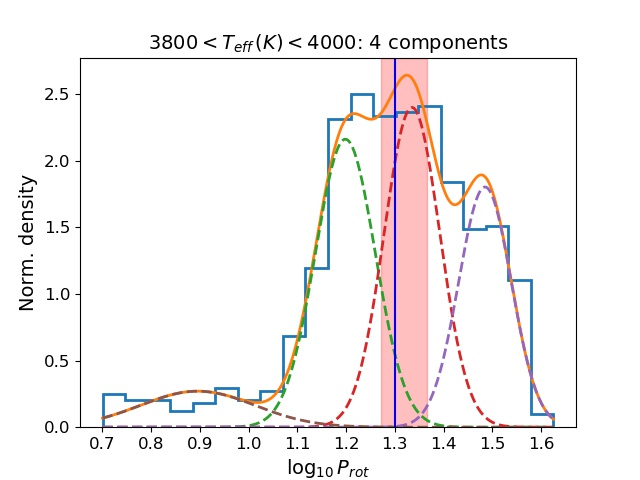}
  \end{minipage}
  \begin{minipage}[b]{0.33\textwidth}
  \includegraphics[width=\textwidth]{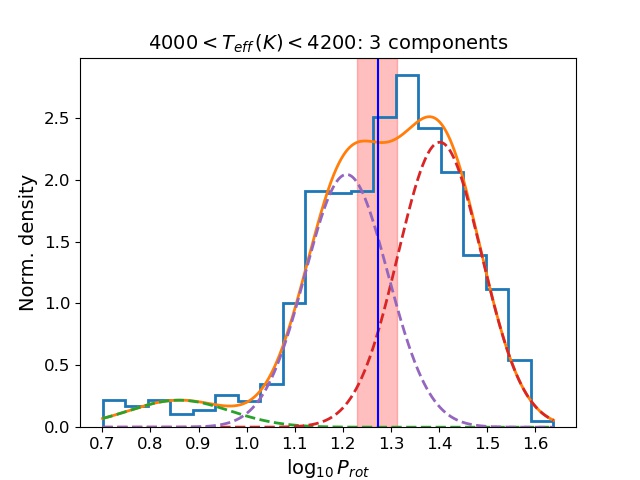}
  \end{minipage}
  \hfill
  \begin{minipage}[b]{0.33\textwidth}
  \includegraphics[width=\textwidth]{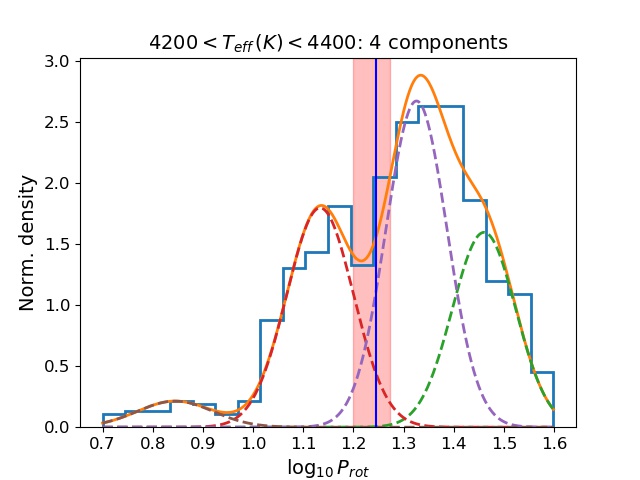}
  \end{minipage}
  \hfill
  \begin{minipage}[b]{0.33\textwidth}
  \includegraphics[width=\textwidth]{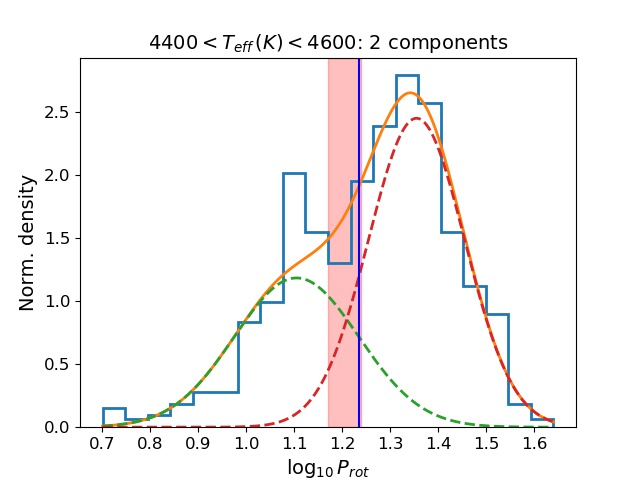}
  \end{minipage}
  \begin{minipage}[b]{0.33\textwidth}
  \includegraphics[width=\textwidth]{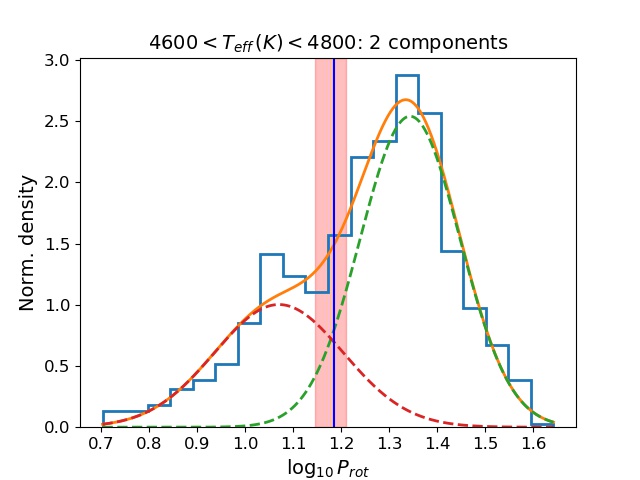}
  \end{minipage}
  \hfill
  \begin{minipage}[b]{0.33\textwidth}
  \includegraphics[width=\textwidth]{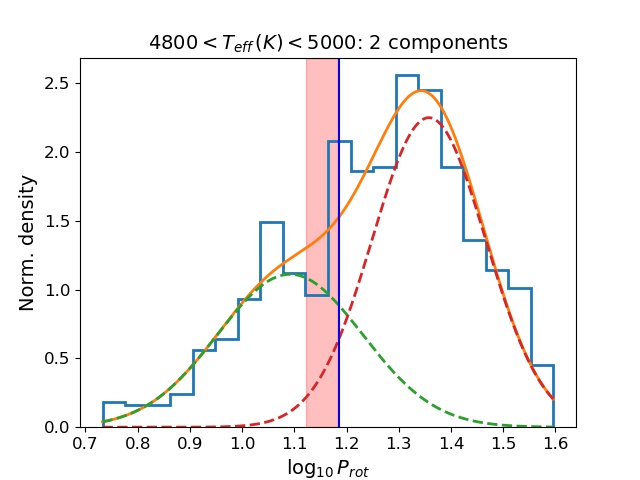}
  \end{minipage}
  \hfill
  \begin{minipage}[b]{0.33\textwidth}
  \includegraphics[width=\textwidth]{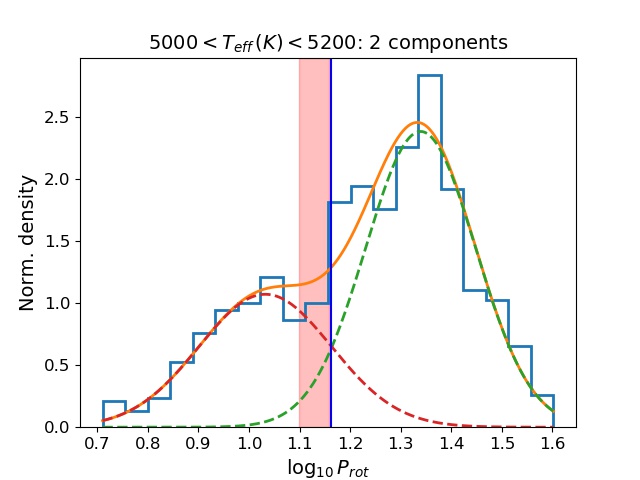}
  \end{minipage}
  \begin{minipage}[b]{0.33\textwidth}
  \includegraphics[width=\textwidth]{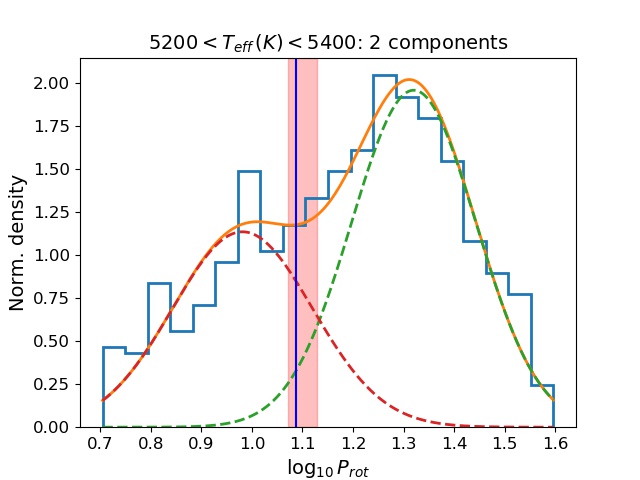}
  \end{minipage}
  \hfill
  \begin{minipage}[b]{0.33\textwidth}
  \includegraphics[width=\textwidth]{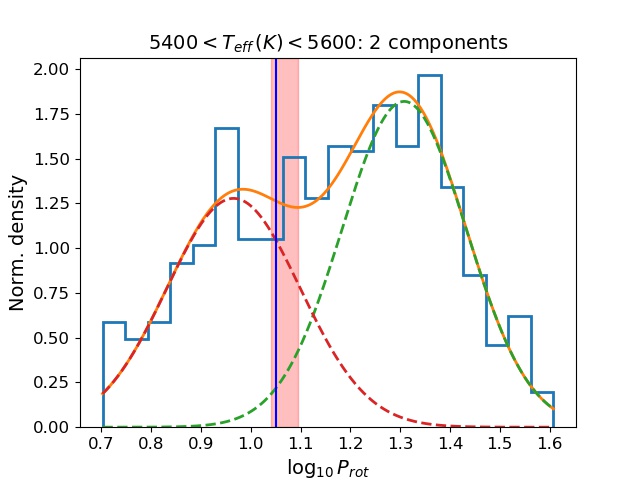}
  \end{minipage}
  \hfill
  \begin{minipage}[b]{0.33\textwidth}
  \includegraphics[width=\textwidth]{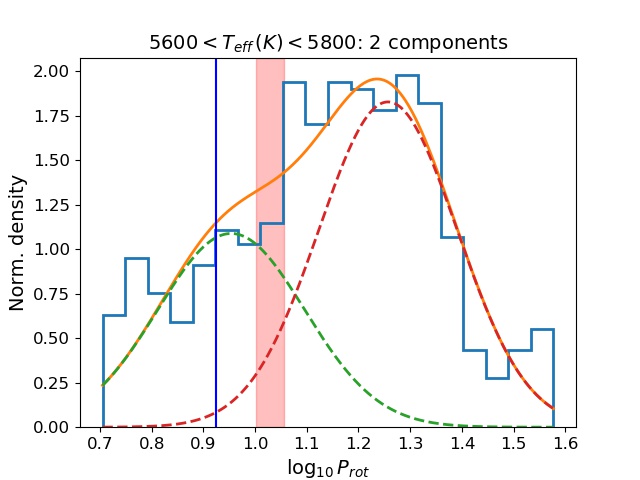}
  \end{minipage}
  \caption{
  Distribution of the logarithm of the rotation periods for different temperature bins (light blue). The Gaussian mixture model is shown as orange curve, and the dashed colored lines show the individual Gaussians. The vertical blue lines indicate the period at the local variability minimum (see Fig.~\ref{Prot_range}), and the shaded red area indicates the expected period range $\Prot \pm \Delta P_{\rm rot}$ for a 750\,Myr isochrone.}
  \label{gmm}
\end{figure*}

In Fig.~\ref{gmm} we show the distribution of the logarithm of the rotation period $\Prot$ for the same temperature bins as in Fig.~\ref{Prot_range}. The data are fit by a Gaussian mixture model, which fits the period distribution by a combination of several Gaussians. The number of Gaussians needed to properly fit the given data is obtained following the Bayesian information criterion (BIC). This number is computed for a combination of up to 20~Gaussians, where the model returning the lowest BIC is preferred. The period distributions in Fig.~\ref{gmm} were fit with 2--4 components, depending on the temperature bin. In all cases that were fit with 3--4~components, at least one Gaussian is used to fit the short periods tail covering $\Prot<10$~days, which is not of interest here. In all temperature bins we find that a bimodal distribution is preferred over a unimodal distribution. Despite the fact that in the temperature range 3800--4200\,K the bimodality is not visible by eye.

As stated above, we expect the dip of the period distribution to coincide with the periods of minimum variability along the 750\,Myr isochrone. 
The period ranges $\Prot\pm\Delta P_{\rm rot}$ from Table~\ref{Prot_uncertainties} are shown as vertical red bars in Fig.~\ref{gmm}, and the periods at the detected variability minima are indicated by vertical blue lines. The vertical red bar is not indicated for the first temperature bin (upper left panel in Fig.~\ref{gmm}) because the predicted period is close to  the edge of our detection threshold of $\sim$40~days. For all other temperatures bins, the period range indicated by the red bar shows remarkably good agreement with the dip of the rotation period distribution. We note that, for the hottest temperature bin (lower right panel in Fig.~\ref{gmm}), the deviation between the red area ($\Prot=10.71\pm0.65$~days) and the blue line ($\Prot=8.41$~days) may still indicate consistency within $2\sigma$ if we assume an uncertainty of 0.65~days also on the period of the local variability minimum.




\section{Discussion}\label{discussion}
This study provides the first comprehensive analysis of rotation periods covering all K2 observing campaigns. Table~\ref{camp_table} shows that the number of stars satisfying all selection criteria varies between $\sim 11-27\%$ among different campaigns.

Measuring accurate rotation periods of slowly-rotating stars is challenging owing to limited spot lifetimes and/or instrumental systematics mimicking rotational variability. As shown in Fig.~\ref{Period12}, depending on the length of the observing campaign, the reliability of the derived periods significantly decreases for stars with periods above $\sim$20~days. We conclude that the dearth region is not that evident in Fig.~\ref{Teff_Prot} because the derived periods are much more uncertain compared to previous rotation periods measurements in \textit{Kepler}. By combining all observing quarters for a certain \textit{Kepler} star, up to $\sim$4~years in total, \citet{McQuillan2014} and \citet{Reinhold2015} could significantly reduce the number of spurious period detections, which is not possible for the majority of the K2 stars observed only for $\sim$80~days (see appendix \ref{method_Kepler}). Additionally, the K2 mission has a reduced photometric precision due to the new mission concept, which further complicates the measurements of small rotational signatures expected for slow rotators with periods between 15--25~days, where the dearth region is expected. 

\subsection{Comparison with literature periods}
\begin{figure}
  \resizebox{\hsize}{!}{\includegraphics{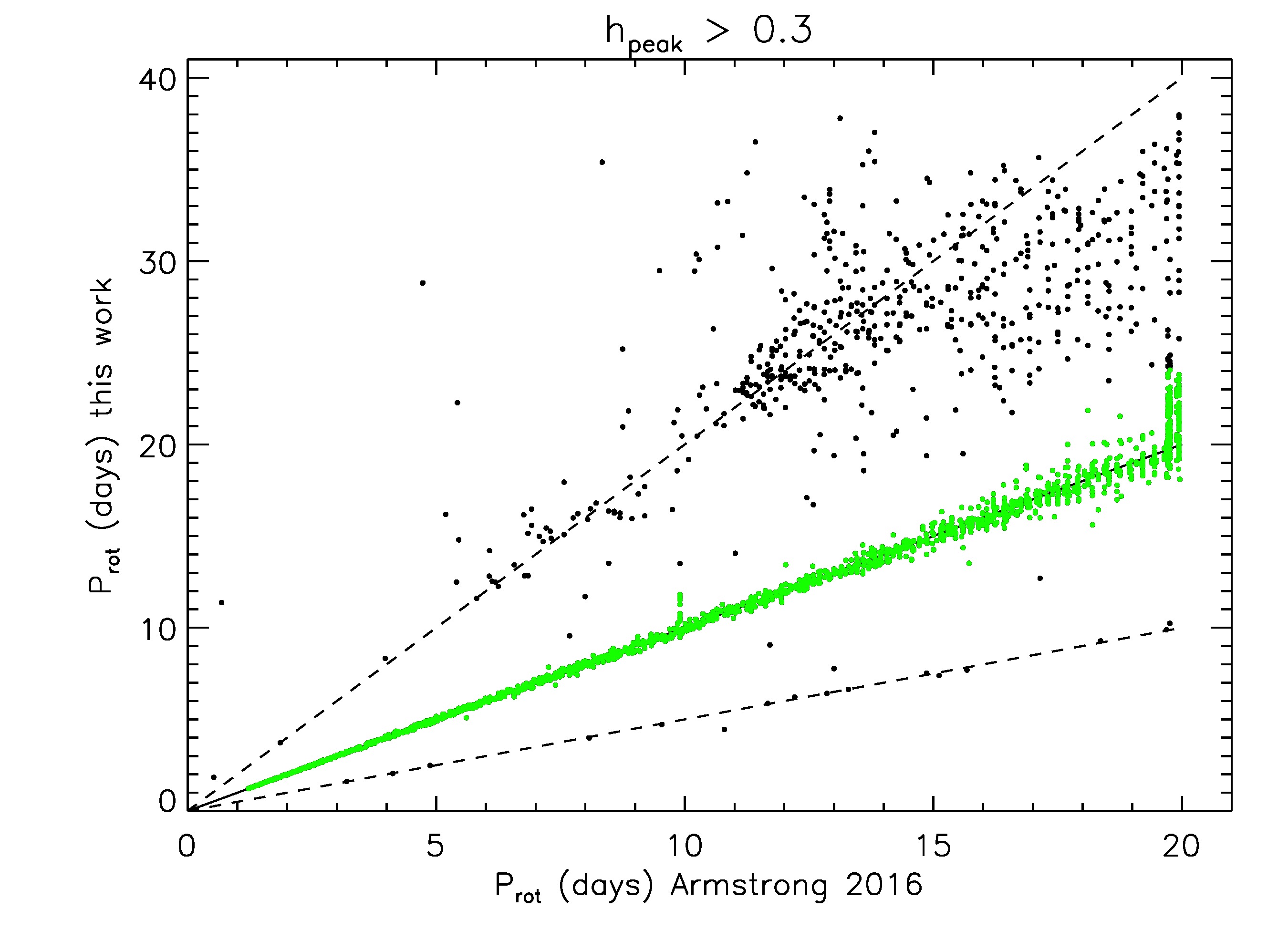}}
  \caption{Comparison of rotation periods derived by \citet{Armstrong2016} to this work. The green dots show periods deviating by less than 20\% among each other. The solid black line shows the 1:1 identity, and the dashed lines show the 1:2 and 2:1 period ratios.}
  \label{periods_Armstrong2016}
\end{figure}

The rotation periods derived in this work show remarkable good agreement with previous K2 rotation period measurements. \citet{Armstrong2016} analyzed the campaigns C0--C4, and found 9400~periods attributed to stellar rotation. Our sample has 2591~stars in common, of which 2011~stars show periods deviating by less than 20\% (green dots in Fig.~\ref{periods_Armstrong2016}). Most deviations arise from the fact that these authors limited their periods to 20~days, so in many cases only the half period is detected (see upper dashed line in Fig.~\ref{periods_Armstrong2016}). 

\begin{figure}
  \resizebox{\hsize}{!}{\includegraphics{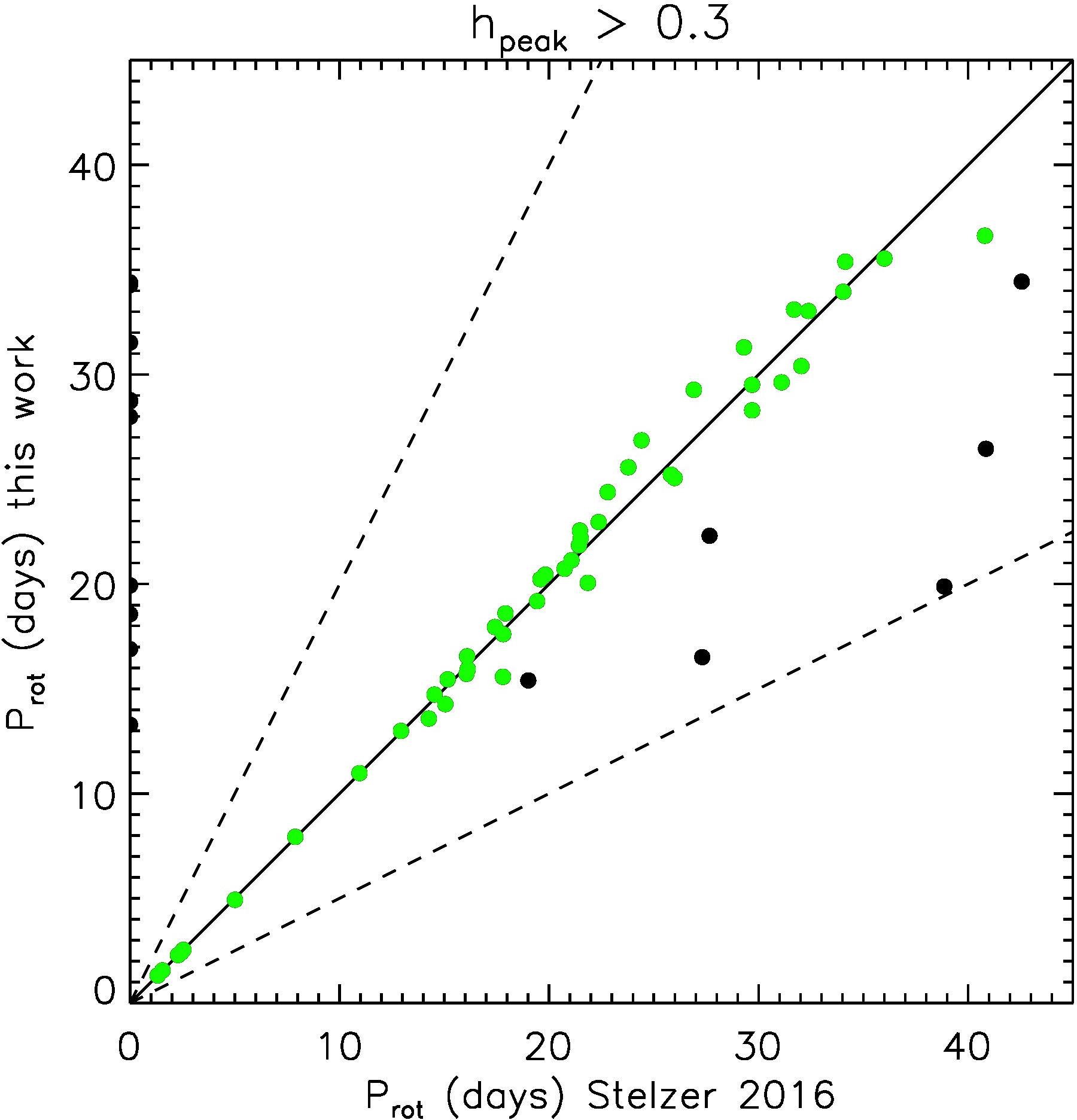}}
  \caption{Comparison of rotation periods derived by \citet{Stelzer2016} to this work. Symbols and lines are the same as in Fig.~\ref{periods_Armstrong2016}.}
  \label{periods_Stelzer2016}
\end{figure}

\citet{Stelzer2016} analyzed 134 M~dwarfs observed in the K2 campaigns C0--C4 (see Fig.~\ref{periods_Stelzer2016}). 65~stars are included in our sample. Thereof, 47~stars exhibit periods deviating by less than 20\%. For 10~stars we detected a period whereas \citet{Stelzer2016} did not report one. Furthermore, these authors found a clear transition of the photometric variability between fast and slow rotators at a period of $\sim$10~days. This result is consistent with the dependence of the variability range on rotation period in the top row in Fig.~\ref{Prot_range}: for all temperature bins lower than 4000\,K, $\Rvar$ shows high values and a rather flat dependence on $\Prot$ for periods less than 10~days, and steeply decreases towards longer periods.

\begin{figure}
  \resizebox{\hsize}{!}{\includegraphics{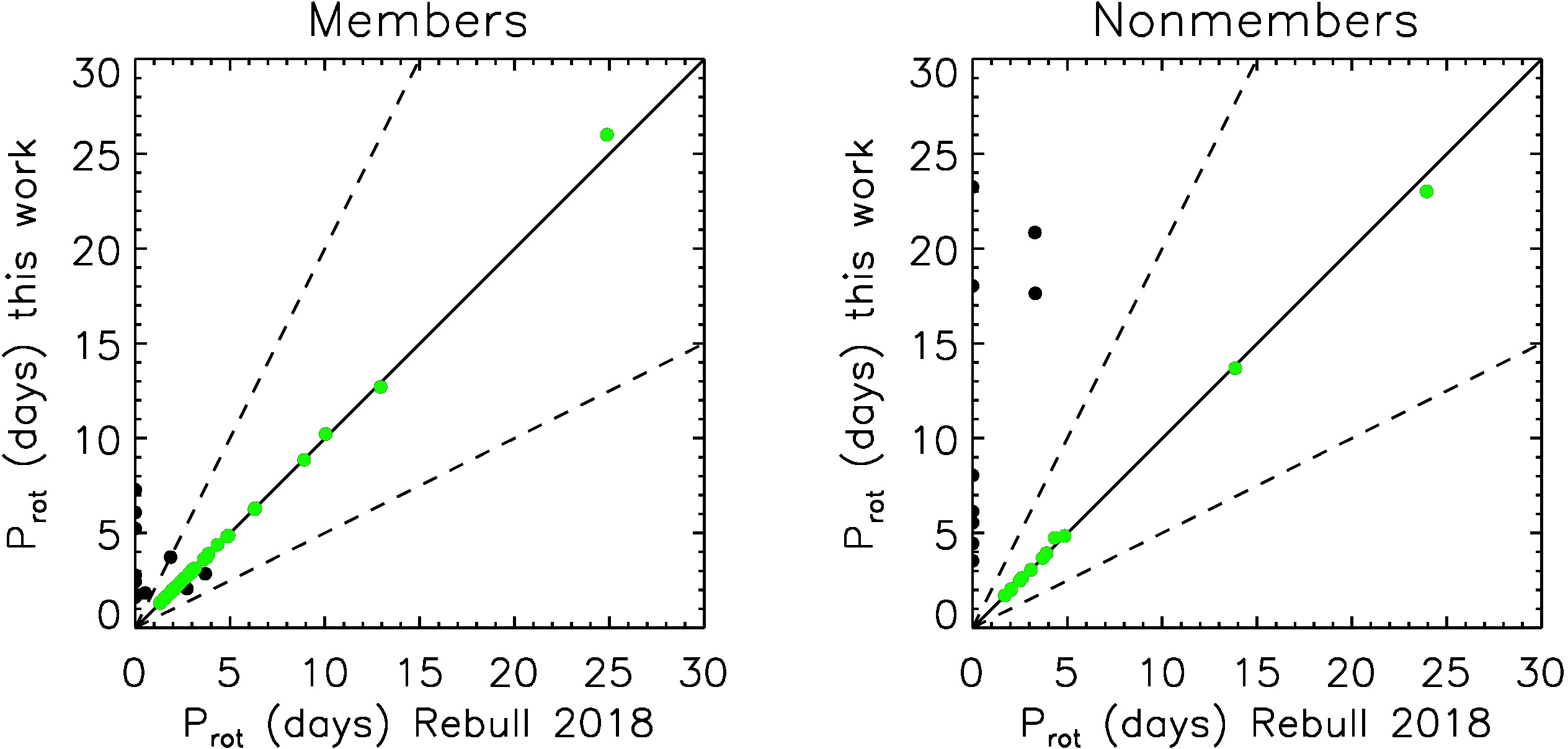}}
  \caption{Comparison of rotation periods derived by \citet{Rebull2018} for members (left) and nonmembers (right) of the Upper Scorpius and $\rho$ Ophiuchus star clusters to this work. Symbols and lines are the same as in Fig.~\ref{periods_Armstrong2016}.}
  \label{periods_Rebull2018}
\end{figure}

In Fig.~\ref{periods_Rebull2018} we compare our measurements to the rotation periods derived by \citet{Rebull2018} for members (left) and nonmembers (right) of the Upper Scorpius and $\rho$ Ophiuchus star clusters. For all stars we find very good agreement among the derived periods with few exceptions. We derive 6~periods for the members and  7~periods for the nonmembers where \citet{Rebull2018} does not report a period.

\begin{figure}
  \resizebox{\hsize}{!}{\includegraphics{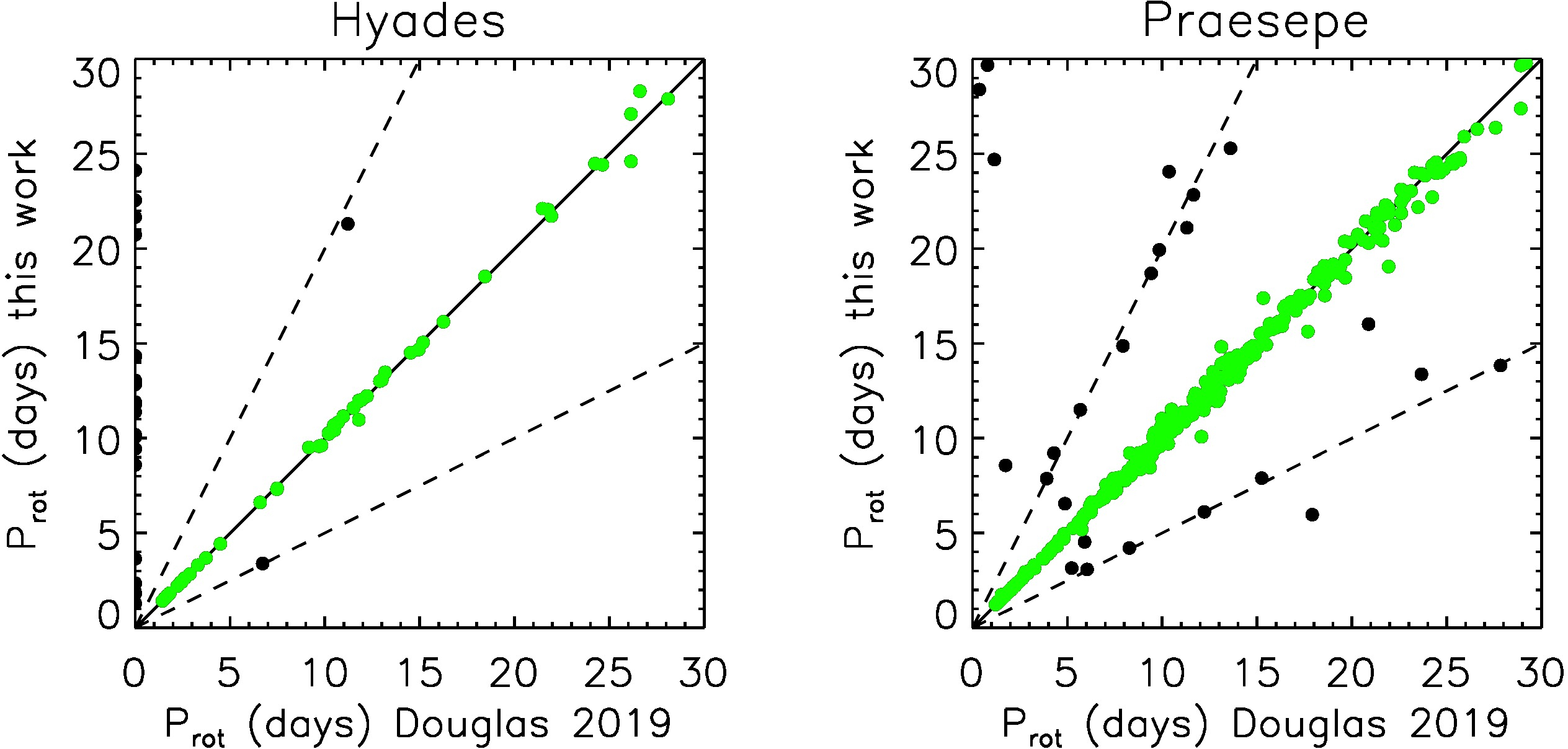}}
  \caption{Comparison of rotation periods derived by \citet{Douglas2019} for the Hyades (left) and Praesepe (right) star clusters to this work. Symbols and lines are the same as in Fig.~\ref{periods_Armstrong2016}.}
  \label{periods_Douglas2019}
\end{figure}

The rotation periods of stars in the Hyades and Praesepe star clusters are compared to our work in Fig.~\ref{periods_Douglas2019}. The periods derived for both clusters show excellent agreement. In the Hyades we find 22~periods that were not reported by \citet{Douglas2019}. For the Praesepe stars we often detect the double period, and sometimes the half period. These cases might be worth checking for a more detailed cluster study.

\subsection{The dearth region: Implications from gyrochronology}
\begin{figure}
  \resizebox{\hsize}{!}{\includegraphics{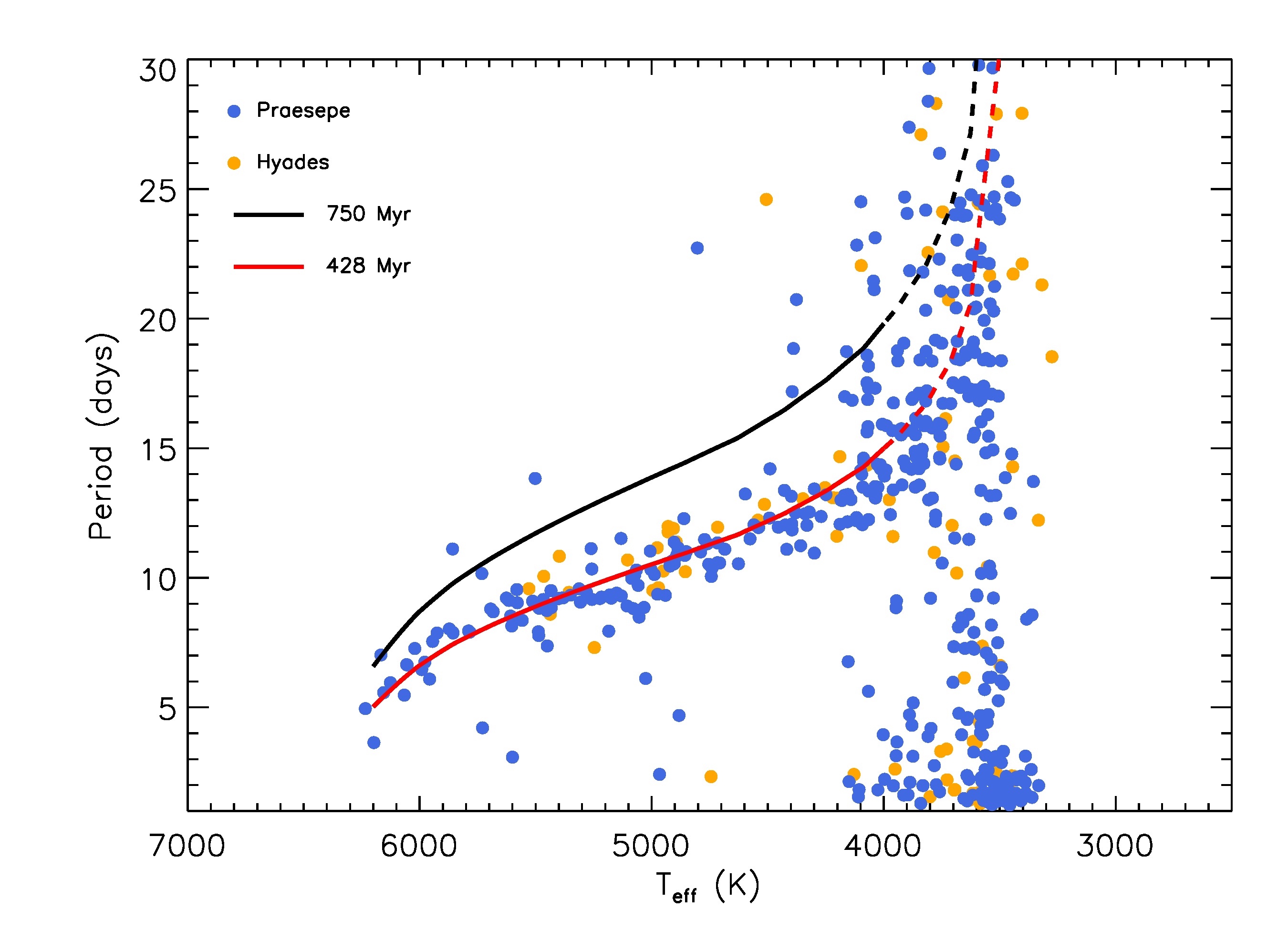}}
  \caption{Rotation periods of the Praesepe (blue) and the Hyades (orange) star clusters derived by \citet{Douglas2019} vs. effective temperature $T_{\rm eff}$. Gyrochronology isochrones of 750\,Myr and 428\,Myr are shown in black and red, respectively. The solid lines down to 4000\,K indicate the temperature range used in the fit, and the dashed lines show the extrapolation of the fit down to 3500\,K.}
  \label{cluster_periods}
\end{figure}

As mentioned earlier, recent studies \citep{Curtis2019,Douglas2019} demonstrated that stars in open clusters spin faster than predicted by gyrochronology at the given cluster age. This finding also affects our results since the stars along the 750\,Myr isochrone in Figs.~\ref{Teff_Prot} and~\ref{Teff_Prot2} may be older than predicted from their rotation periods. To test this hypothesis, we consider the periods derived by \citet{Douglas2019} for the Hyades and Praesepe clusters. In Fig.~\ref{cluster_periods} we combine these period measurements since both clusters are considered as almost coeval (i.e. 600--800\,Myr), and plot them against the effective temperatures of the stars. We fit the slow-rotator sequence of the cluster stars hotter than 4000\,K (solid lines) in the period-temperature plane using the relation of \citet{Barnes2010}, with the age as the only free parameter. The best fit gyrochronology age of 428\,Myr (red line) comes out younger than the commonly considered age of 600--800\,Myr for these clusters. For comparison, an isochrone of 750\,Myr is over-plotted (black line), which lies well above the cluster periods. This result allows different conclusions: 1) The cluster stars spin faster than predicted from gyrochronology because rotational braking ceases or becomes very inefficient (for yet unknown reason) in these clusters. 2) The cluster stars might actually be younger than thought so far -- assuming that gyrochronology relations are correct! This result supports the hypothesis that the stars at the local variability minimum may be older than 750\,Myr. 
A more detailed analysis of periods in different clusters would be required to properly address this problem. We leave this analysis to a future publication.

The rotation period bimodality seen in the \textit{Kepler} field (and in other fields as shown in this study) might also be explained by an epoch of increased spin-down efficacy. During that epoch stars would spin down much faster, and would quickly be moved over the dearth region. This explanation has already been proposed by \citet{McQuillan2014}. In the light of the observation of stalled spin-down in the open clusters, i.e. the opposite behavior, this scenario appears unlikely, unless the magnetic field topology (at some age) reconfigures such that magnetic braking becomes much more efficient again.

\section{Conclusions}
We showed that the joint rotation period distribution of the selected K2 sample exhibits a high degree of conformity with the rotation period distribution of stars in the \textit{Kepler} field. Both distributions show a lack of intermediate rotators (i.e. a dearth region), accompanied by a decrease of variability. This result supports the explanation proposed by \citet{Reinhold2019} that the dearth region reflects a non-detection of periodicity owing to a variability decrease below common detection thresholds in the period retrieval. To add a word of caution: the proposed effect of spot and faculae cancelation might not be that severe such that rotation periods cannot be detected at all. Actually, we have detected many rotation periods in the range 15--25~days, showing a large spread in variability (see Fig.~\ref{Prot_range}). However, a temporary increase of faculae (or decrease of spot) filling factors can explain the local minimum of variability at certain rotation periods. At what age this scenario might occur seems less clear given that gyrochronology might be a less reliable age-dating method than previously thought.

The proposed explanation for the dearth region of two stellar populations of different age in the \textit{Kepler} field becomes less likely in the light of our results. Given the fact that the observed K2 fields do not overlap with the \textit{Kepler} field, our results strongly suggest that the lack of stars at intermediate rotation periods is independent of the field of view, and represents a general manifestation of stellar activity. Our results further lead us to conclude that the dependence of the photometric variability on the rotation period is more complex than previously thought.

Furthermore, the interplay of bright regions with dark spots became increasingly important in recent years. \citet{Rackham2018} studied the influence of unocculted spots and faculae on the derived parameters of transiting exoplanets. \citet{Montet2017} showed that stellar variability on activity cycle time scales is either correlated or anti-correlated with the short-term variability caused by star spots, with a transition from spot-dominated to faculae-dominated activity at periods between 15--25~days. This period range is consistent with the age of 2550\,Myr derived by \citet{Reinhold2019} for this transition. \citet{Morris2018} reported the potential discovery of bright star spots on the planet-hosting star TRAPPIST-1.
Ongoing missions such as TESS will provide a new view on the complexity of the activity-rotation relation in the near future.

\begin{acknowledgements}
We like to thank the referee for providing insightful comments that greatly helped to improve the manuscript. The research leading to the presented results has received funding from the European Research Council under the European Community's Seventh Framework Programme (/2007-2013) / ERC grant agreement no 338251 (StellarAges), and partly from the European Research Council (ERC) under the European Union's Horizon 2020 research and innovation programme (grant agreement No. 715947). TR would like to thank the International Space Science Institute, Bern, for their support of science team 446 and the resulting helpful discussions.
\end{acknowledgements}
\bibliographystyle{aa}
\bibliography{biblothek}

\begin{appendix}
\section{K2 data products}\label{app_A}
\begin{figure*}
  \centering
  \includegraphics[width=17cm]{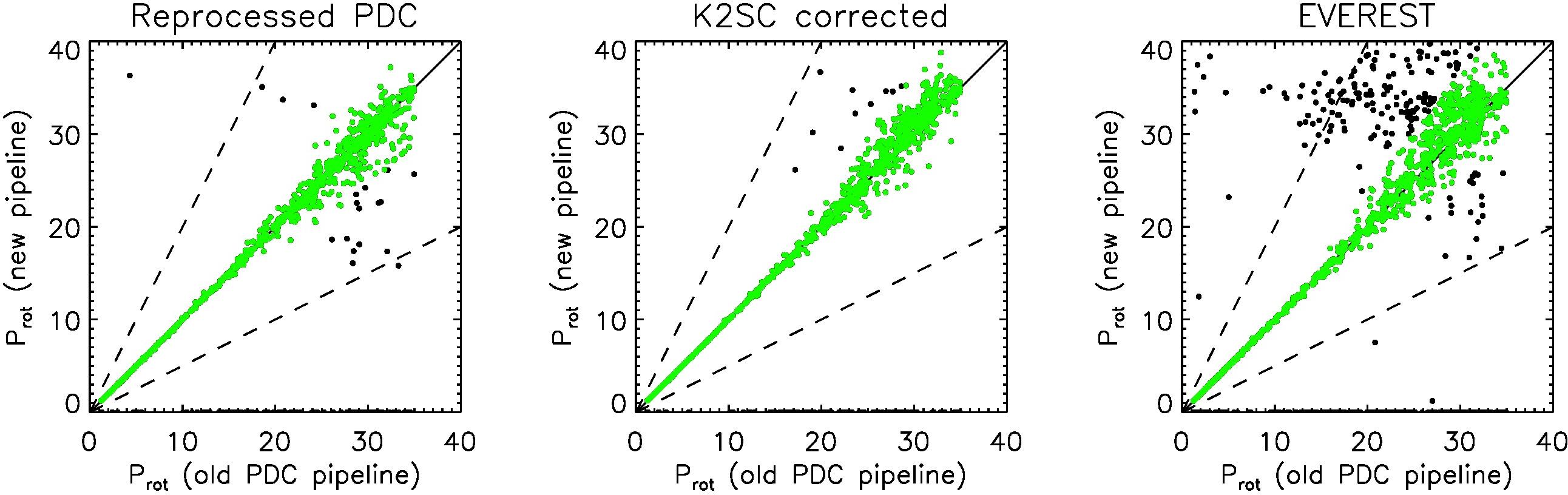}
  \caption{Period comparison between different pipelines. Left panel: reprocessed PDC data. Middle panel: K2SC reduced data. Right panel: EVEREST data reduction. Symbols and lines are the same as in Fig.~\ref{periods_Armstrong2016}-\ref{periods_Douglas2019}.}
  \label{periods_camp4_comparison_crop}
\end{figure*}

K2 data contains many instrumental systematics due to the loss of \textit{Kepler's} third reaction wheel\footnote{For details see, e.g., https://keplerscience.arc.nasa.gov/K2/MissionConcept.shtml}.
Different pipelines have been developed and tested to remove the instrumental signals from the data while preserving as much astrophysical signal as possible. To test the robustness of our results against different data reductions, we re-analyzed all stars in campaign~4 that have been reduced with different pipeline versions. The first one is an updated version of the PDC pipeline\footnote{For further information, see https://keplerscience.arc.nasa.gov/k2-uniform-global-reprocessing-underway.html}, the second one is the K2SC pipeline \citep{Aigrain2016}, and the third one is the EVEREST pipeline (see \citealt{Luger2016} for version~1.0 and \citealt{Luger2018} for version~2.0 used here). In Fig.~\ref{periods_camp4_comparison_crop} we show the periods derived by the old PDC pipeline vs. the new pipeline for the reprocessed (left panel), the K2SC (middle panel), and the EVEREST (right panel) data. Our measurements show remarkably good agreement among the different pipelines. For the reprocessed and the K2SC data, almost 93\% of all stars (green dots) show deviations less than 20\% between the derived periods. In 102 (reprocessed) and 106 (K2SC) cases, the new pipeline did not satisfy the required criteria, such that the periods were set to zero. For the EVEREST pipeline, however, we find that roughly 76\% of the periods agree within 20\%. This number is still considered as good agreement, although the percentage is lower than for the other two data sets. This deviation relies on the fact that no periodicity was found in 221~cases  satisfying the imposed criteria. When only considering cases where periods were detected in both pipelines, the agreement among them exceeds 91\%. We conclude that the results presented in this study do not depend on the choice of the data reduction pipeline, as long as stellar variability is preserved in the processing.

\section{Testing method on \textit{Kepler} data}\label{method_Kepler}
\begin{figure}
  \resizebox{\hsize}{!}{\includegraphics{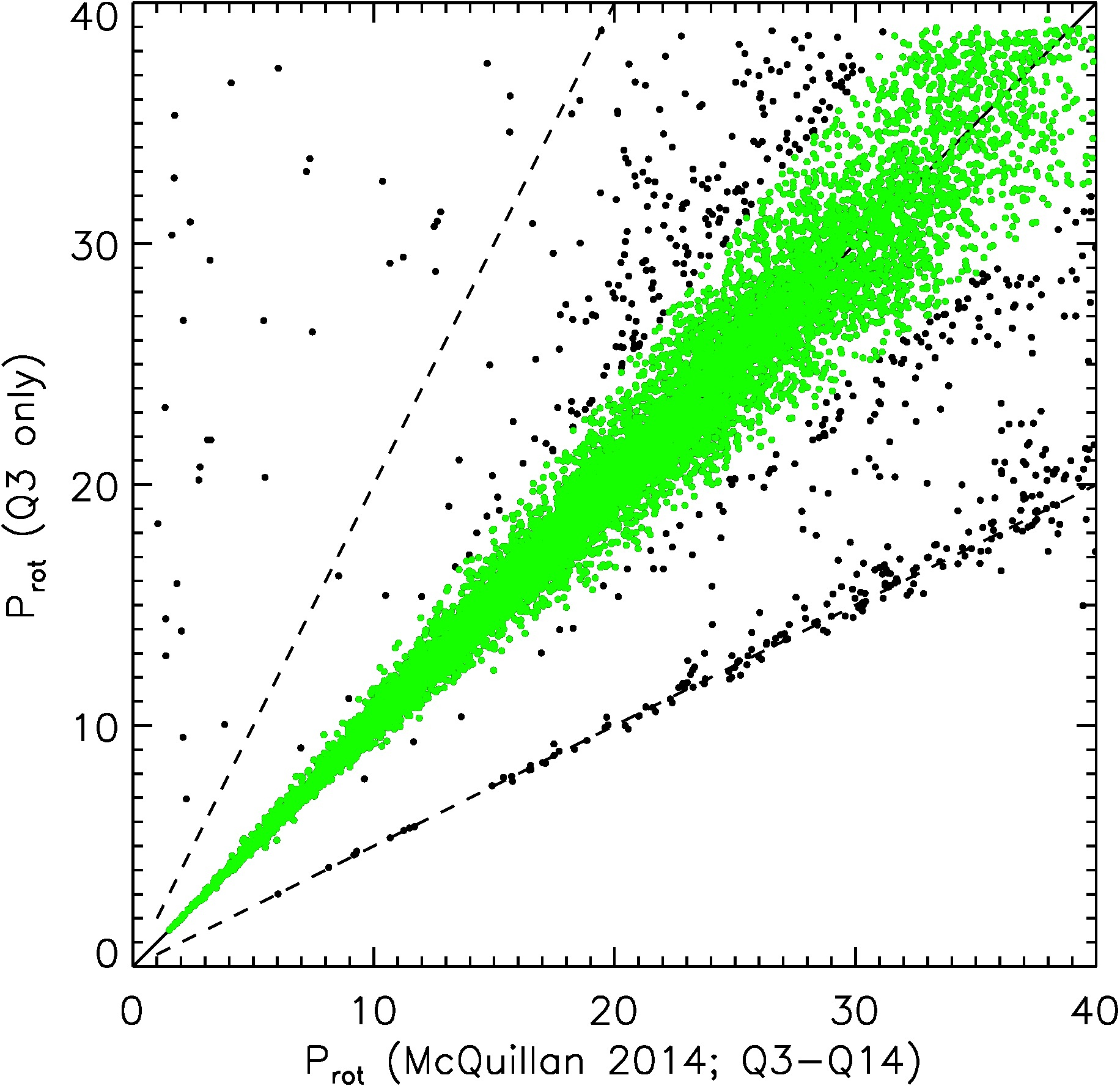}}
  \caption{Period comparison between \textit{Kepler} Q3 data, and periods derived by \citet{McQuillan2014} using Q3-Q14 data. Symbols and lines are the same as in Fig.~\ref{periods_Armstrong2016}-\ref{periods_Douglas2019}.}
  \label{periods_kepler}
\end{figure}

To test the reliability of our period detection criteria, we applied the exact same analysis to \textit{Kepler} quarter~3 (Q3) data. The observing time span of Q3 data is similar to the K2 observing campaigns. Our analysis reveals rotation periods for 25,675~stars in the full Q3 data set of 125,292~stars in the considered parameter range (i.e. roughly 20.5\%). This number is comparable to the number of stars selected from each K2 campaign (see Table~1). In Fig.~\ref{periods_kepler} we compare the periods derived using only Q3 data, to the periods obtained by \citet{McQuillan2014} using Q3--Q14 data. The latter are considered as valid references since the underlying observing time span is much longer such that periodicity will be picked up more easily, especially for slow rotators. \citet{McQuillan2014} derived rotation periods for 14,854 of the 25,675~stars. We find that 14,247 of the 14,854 of the periods (almost 96\%) agree within 20\%, which we consider as an excellent agreement (green dots).

We derive period uncertainties by comparing the periods from Fig.~\ref{periods_kepler}, equivalent to Fig.~\ref{delta_Period12}. The period uncertainties derived for the \textit{Kepler} stars are smaller than those obtained from the comparison of two different K2 campaigns. This results from the fact that the \textit{Kepler} periods are more reliable because they were obtained from a much longer time series. The exponential fit to the data points (red curve) should be interpreted as minimum rotation period uncertainty and was over-plotted in Fig.~\ref{delta_Period12} for comparison.

\begin{figure}
  \resizebox{\hsize}{!}{\includegraphics{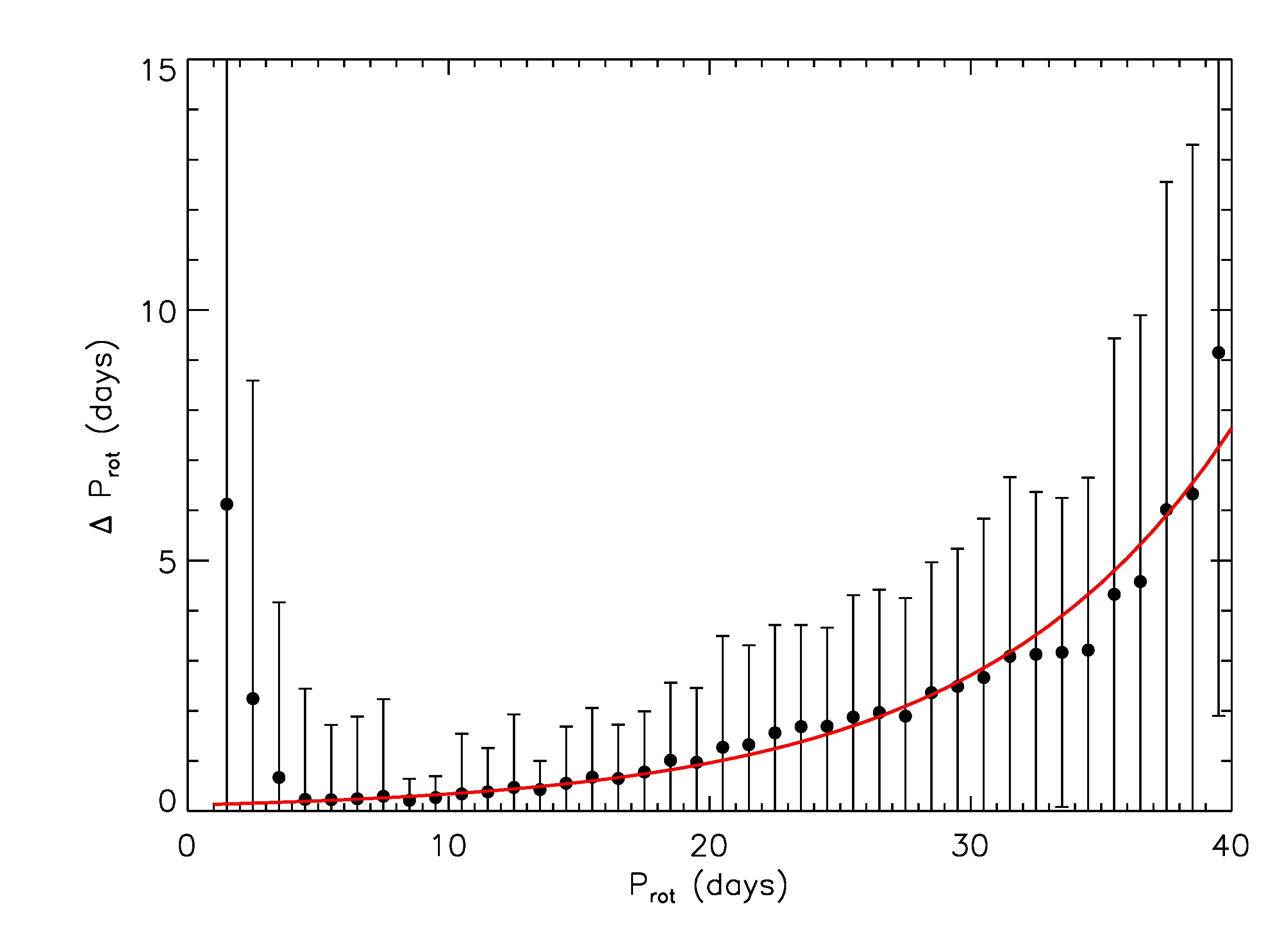}}
  \caption{Estimate of the rotation period uncertainty (as described in Sect.~\ref{periods}) comparing periods derived from \textit{Kepler} Q3 data to those obtained from the analysis of the full light curve. The red curve shows an exponential fit to the (equally weighted) data points. This red curve is shown for comparison in Fig.~\ref{delta_Period12}.}
  \label{uncertainties_kepler}
\end{figure}

\end{appendix}

\end{document}